\documentclass{aa}
\usepackage{amssymb,amsmath}
\usepackage{graphicx}
\usepackage[varg]{txfonts}
\usepackage[colorlinks=true, citecolor=blue, linkcolor=blue, urlcolor=blue]{hyperref}
\usepackage{natbib}
\bibpunct{(}{)}{;}{a}{}{,} 

\newcommand\Tstrut{\rule{0pt}{2.6ex}}         

\begin{document}

\title{Probing spatial homogeneity with LTB models: a detailed discussion}

\titlerunning{Probing spatial homogeneity with LTB models}

\author{
M. Redlich \inst{1,2}
\and
K. Bolejko \inst{2}
\and
S. Meyer \inst{1}
\and
G. F. Lewis \inst{2}
\and
M. Bartelmann \inst{1}
}

\authorrunning{M. Redlich et al.}

\institute{
Zentrum f\"ur Astronomie der Universit\"at Heidelberg, Institut f\"ur 
Theoretische Astrophysik, Albert-Ueberle-Str.~2, 69120 Heidelberg, Germany \\
\email{matthias.redlich@stud.uni-heidelberg.de}
\and
Sydney Institute for Astronomy, School of Physics, A28, The University of Sydney, NSW 2006, Australia
}

\date{\emph{A\&A manuscript, version \today}}

\abstract{Do current observational data confirm the assumptions of the cosmological principle, or is there statistical evidence for deviations from spatial homogeneity on large scales? To address these questions, we developed a flexible framework based on spherically symmetric, but radially inhomogeneous Lema{\^\i}tre-Tolman-Bondi (LTB) models with synchronous Big Bang. We expanded the (local) matter density profile in terms of flexible interpolation schemes and orthonormal polynomials. A Monte Carlo technique in combination with recent observational data was used to systematically vary the shape of these profiles. In the first part of this article, we reconsider giant LTB voids without dark energy to investigate whether extremely fine-tuned mass profiles can reconcile these models with current data. While the local Hubble rate and supernovae can easily be fitted without dark energy, however, model-independent constraints from the Planck 2013 data require an unrealistically low local Hubble rate, which is strongly inconsistent with the observed value; this result agrees well with previous studies. In the second part, we explain why it seems natural to extend our framework by a non-zero cosmological constant, which then allows us to perform general tests of the cosmological principle. Moreover, these extended models facilitate explorating whether fluctuations in the local matter density profile might potentially alleviate the tension between local and global measurements of the Hubble rate, as derived from Cepheid-calibrated type Ia supernovae and CMB experiments, respectively. We show that current data provide no evidence for deviations from spatial homogeneity on large scales. More accurate constraints are required to ultimately confirm the validity of the cosmological principle,
however.} 


\maketitle

\section{Introduction}

While the discovery of the (apparently) accelerated expansion of the Universe established a non-vanishing dark energy contribution in the framework of the standard cosmological model \citep{1998AJ....116.1009R,1999ApJ...517..565P}, these observations also motivated many researchers to question the theoretical foundations the standard model is built upon. One of these foundations is the cosmological principle, which asserts that our Universe is spatially isotropic and homogeneous when averaged over sufficiently large scales ($\gtrsim 100 \, \mathrm{Mpc}$). This assumption is truly remarkable because  if it is correct, it implies that the large-scale geometry of our Universe -- which is notably inhomogeneous on small scales -- is accurately described by the simple class of spatially isotropic and homogeneous Friedman-Lema{\^\i}tre-Robertson-Walker (FLRW) models \citep{1935ApJ....82..284R, 1935QJMat...6...81W}. Statistical isotropy about our position has been confirmed by the remarkable uniformity of the cosmic microwave background (CMB) spectrum \citep{2013ApJS..208...20B, 2013arXiv1303.5076P}. In contrast, statistical homogeneity on large scales ($\sim$ Gpc scales) is hard to confirm, mainly because it is difficult to distinguish a temporal from a spatial evolution on the past light cone \citep[see][for reviews]{2011RSPTA.369.5115M, 2012CRPhy..13..682C}. 

This uncertainty inspired many to study inhomogeneous cosmologies \citep[see][for comprehensive reviews]{2011CQGra..28p4004M, 2011CQGra..28p4002B}, including non-Copernican models that explain the apparent accelerated expansion of the Universe by means of radial inhomogeneities without requiring any form of dark energy. The basic idea behind these alternative models is quite simple because we know from observations and numerical simulations that the large-scale structure of the Universe consists of filaments and voids \citep[see e.g.][]{2005ApJ...624...54H, 2005Natur.435..629S, 2010AIPC.1241..981L, 2011ApJ...736...51E, 2011A&A...532A...5E, labini2011inhomogeneities, 2012MNRAS.425..116S, 2013MNRAS.429.2910C, 2013MNRAS.434..398N, 2014MNRAS.440.1248N, 2014MNRAS.442.3127S, 2014MNRAS.440.2922M}. Einstein's General theory of Relativity tells us that the expansion rate in space-time regions with lower matter density should be higher than in regions with a higher matter density. If we were to live in a large-scale under-density, the local expansion rate around us would be higher than the average expansion rate in the background. Light-rays propagating from distant sources to us -- the observer -- would therefore feel an accelerated expansion rate along their path. In comparison to the standard cosmological model, these scenarios hence replace a cosmic acceleration in time (due to dark energy) by a spatially varying expansion rate.

One particular, exact inhomogeneous cosmological model that has extensively been studied is the Lema{\^\i}tre-Tolman-Bondi (LTB) model, which is a spherically symmetric, but radially inhomogeneous dust solution of Einstein's field equations \citep{1933ASSB...53...51L, 1934PNAS...20..169T, 1947MNRAS.107..410B}. The spatial hypersections of LTB models are spherically symmetric only about one point, and to conserve the remarkable uniformity of the CMB spectrum, we would have to live very close to the symmetry centre \citep{2006PhRvD..74j3520A, 2010PhRvD..82j3532F}. Interpreted as a faithful representation of the Universe, these void models breach the Copernican principle and require a tremendous fine-tuning of our position in the Universe. The plausibility of such scenarios is therefore more than dubious \citep[see][however, for interesting thoughts]{2012A&A...543A..71C}. The standard cosmological model also requires significant fine-tuning, however, which gives rise to controversial philosophical discussions \citep[see e.g.][]{2008GReGr..40..301D, 2010deto.book.....A}. In this work, we simply demonstrate that a quite general class of LTB void models without dark energy is inconsistent with current observational data, which allows us to set the philosophical discussion aside.

The vast literature on inhomogeneous cosmologies -- in particular LTB models -- is summarised in the review articles by \citet{2011CQGra..28p4002B} and \citet{2011CQGra..28p4004M}, which allows us to only focus on the aspects that are particularly relevant for this work. For reasons to be clarified later, we only discuss LTB models with synchronous Big Bang throughout, meaning that the bang time function is constant and the Universe has the same global age everywhere. In the first part of this work, we additionally set the cosmological constant to zero. It has long been known that LTB models without dark energy can easily fit supernovae, explaining the apparent acceleration of the Universe by a Gpc-scale void around us \citep{2000A&A...353...63C}. In addition, these models can be tuned to fit the small-angle CMB spectrum \citep{2008PhRvL.101y1303Z}. However, most recent studies agree that a good fit to the CMB requires an unrealistically low local Hubble rate of ${H_0 \lesssim 60 \, \mathrm{km} \, \mathrm{s}^{-1} \, \mathrm{Mpc}^{-1}}$, which is strongly inconsistent with the observed value of ${H_0 = (73.8 \pm 2.4) \, \mathrm{km} \, \mathrm{s}^{-1} \, \mathrm{Mpc}^{-1}}$  measured with Cepheid-calibrated type Ia supernovae \citep{2010JCAP...11..030B, 2011PhRvD..83j3515M, 2011ApJ...730..119R, 2012PhRvD..85b4002B, 2012JCAP...10..009Z}. We here already anticipate that we shall finally arrive at the very same conclusion. 

Nevertheless, we believe that our work meaningfully complements the current literature mainly for the following reasons: For the first time, we compare LTB models with the latest Planck 2013 data \citep{2013arXiv1303.5076P}. This is interesting because the Planck data favour a lower Hubble rate than previous CMB experiments. Moreover, we advocate the use of a recently developed technique for analysing CMB spectra in a model-independent manner, which is particularly useful for investigating alternative cosmological models. Secondly, most recent studies have assumed certain functional forms for the mass or curvature profile of LTB models that may be considered characteristic for voids. These empirical parametrisations might simply be too restrictive, however, and certainly impose artificial constraints on the models when performing maximum-likelihood estimates. To our knowledge, only \citet{2008PhRvL.101y1303Z} and \citet{2011PhRvD..83j3515M} considered more flexible spline interpolations for the mass profile of LTB voids. 

We extend these ideas and introduce alternative, flexible parametrisations of the local matter density profile, aiming to impose as little a priori constraints on the detailed form as possible. A Monte Carlo technique in combination with recent observational data allows us to systematically vary the matter density profiles of LTB models and derive statistical constraints on the favoured profile shapes. Moreover, we demonstrate that even the enormous flexibility of radially fine-tuned models does not suffice to simultaneously fit the observed local Hubble rate and the CMB.

After this detailed discussion, we provide simple theoretical arguments that explain why not even heavily fine-tuned LTB models without dark energy can be reconciled with current observational data. We then discuss why it seems most natural to extend our models by a non-zero cosmological constant. The resulting $\Lambda$LTB models constitute a powerful framework for conducting general tests of the cosmological principle \citep{2010JCAP...12..021M}. Recently, \citet{2014MNRAS.438L...6V} proposed a new technique, based on $\Lambda$LTB models, for placing constraints on violations of the Copernican principle. Furthermore, by marginalising over all possible inhomogeneities, these authors derived first observational constraints on the cosmological constant that are free of the usual homogeneity prior \citep[see also][for a similar ansatz]{2013MNRAS.431.1891M}. \citet{2013PhRvL.110x1305M} used $\Lambda$LTB models to investigate whether fluctuations in the local matter density profile can alleviate the well-known discrepancy between the high local Hubble rate as measured by \citet{2011ApJ...730..119R} and the lower one derived from the Planck 2013 data. Obviously, our previously developed, flexible parametrisations of the local matter density profile in combination with a Monte Carlo technique constitute an ideal tool for conducting similar studies. Again, by imposing (almost) no a priori constraints on the detailed shape of the density profiles, our approach is more general than previous works on this field.

The structure of this paper is as follows: In Sect.~\ref{sec:LTB_metric}, we explain our general ansatz for the metric of a Universe that may radially be inhomogeneous on large scales. Section~\ref{sec:obs_data} describes the observational data used to constrain LTB models and also discusses some ambiguities that have to be taken into account when fitting non-standard cosmological models to these data. In Sect.~\ref{sec:LTB_statistical_approach}, we introduce flexible parametrisations of the local matter density profile and describe the algorithm that allows us to statistically constrain the shape of the profile functions. Section~\ref{sec:results_ltb_zero_lambda} summarises our main results of a long series of tests, comparing numerous LTB models without cosmological constant with different combinations of observational data. In Sect.~\ref{sec:theoretical_explanation}, we provide some simple theoretical arguments that explain our empirical results from Sect.~\ref{sec:results_ltb_zero_lambda}. We then extend our models by a non-zero cosmological constant and discuss general probes of the cosmological principle in Sect.~\ref{sec:results_lltb}. Finally, we present our conclusions in Sect.~\ref{sec:conclusions}.

\section{LTB ansatz for the metric}
\label{sec:LTB_metric}

We make the following simplified ansatz for the metric of the observable Universe: We maintain the standard inflationary paradigm and assume that the early Universe was highly homogeneous at least until the time of decoupling. Given the remarkable uniformity of the CMB spectrum, we assume that spherical symmetry about our position was conserved until the present epoch. However, we breach the cosmological principle by allowing a fine-tuned, radially inhomogeneous matter density profile on large scales. As an example, we could envisage to live at the centre of a Gpc-scale void that emerged from an -- admittedly extreme -- under-density of the primordial matter distribution.

Mathematically, this can be realised by describing the local Universe around us with a spherically symmetric, but radially inhomogeneous LTB model that is asymptotically embedded in a homogeneous FLRW background. The LTB model is a spherically symmetric dust solution of Einstein's field equations with zero pressure ($p = 0$), that is, the stress-energy tensor is $T^{\mu \nu} = \rho u^{\mu} u^{\nu}$. This should be an excellent approximation because the local Universe we observe is also a Universe at a late evolutionary epoch, and at this stage, pressure should be completely negligible for describing the large-scale matter distribution. Following the findings of previous work \citep{2010JCAP...11..030B, 2012JCAP...10..009Z}, we generally embed the LTB model into an FLRW background with non-zero curvature to improve the fit to the CMB.

LTB models have already been covered at length in the literature \citep[see e.g.][for many details]{2006igrc.book.....P, 2008GReGr..40..451E, 2011CQGra..28p4002B}. We therefore only briefly summarise the most important equations here. In coordinates comoving with dust and in the synchronous time gauge, the LTB metric can be written as (with $c = 1$)

\begin{equation}
\mathrm{d}s^2 = -\mathrm{d}t^2 + \frac{R'^2(t,r)}{1+2E(r)} \, \mathrm{d}r^2 + R^2(t,r) \, \mathrm{d} \Omega^2 \; ,
\label{eq:LTB_metric}
\end{equation}
where $R(t,r)$ is the areal radius function, a prime denotes the derivative with respect to the radial coordinate $r$ and $E(r)$ is the curvature function. The time evolution of the areal radius function follows from Einstein's field equations and reads

\begin{equation}
\dot{R}^2(t,r) = \frac{2M(r)}{R(t,r)} + 2 E(r) + \frac{\Lambda}{3} \, R^2(t,r) \; ,
\label{eq:LTB_EFE1}
\end{equation}
where a dot indicates the derivative with respect to time, $M(r)$ is the active gravitational mass inside a spherical shell at radius $r$, and $\Lambda$ is the cosmological constant. The second Einstein field equation relates the effective gravitational mass $M(r)$ to the local matter density $\rho(t,r)$,

\begin{equation}
\frac {M'(r)}{R^2(t,r) R'(t,r)} = 4 \pi G \, \rho(t,r) \; .
\label{eq:LTB_EFE2}
\end{equation}
Equation \eqref{eq:LTB_EFE2} will be key for our algorithm in Sect.~\ref{sec:LTB_algorithm}, because we parametrise the local matter density at the present time with flexible basis functions and use Eq. \eqref{eq:LTB_EFE2} to derive the corresponding LTB metric. 

After a separation of variables, Eq. \eqref{eq:LTB_EFE1} can be integrated in time,
\begin{equation}
t_0 - t_{\mathrm{B}}(r) = \int\limits_0^{R(t_0,r)} \frac{1}{\sqrt{\frac{2M(r)}{\tilde{R}} + 2E(r) + \frac{\Lambda}{3} \, \tilde{R}^2(t,r)}} \, \mathrm{d}\tilde{R} \; ,
\label{eq:LTB_age}
\end{equation}
where $t_0$ denotes the present time and $t_{\mathrm{B}}(r)$ is the so-called bang time function. In general, $t_{\mathrm{B}}(r)$ can be an arbitrary function of $r$, which means that the Big Bang does not need to occur synchronously, as in FLRW models. It can be shown, however, that fluctuations in the bang time function can be identified with decaying modes in linear perturbation theory \citep{1977A&A....59...53S, 2008PhRvD..78d3504Z}. Thus, going back in time, these decaying modes would correspond to inhomogeneities at early times. To conserve the remarkable homogeneity of the CMB spectrum, fluctuations in the age of the Universe must have been smaller than a few hundred years \citep{2009GReGr..41.1737B}, which is substantially smaller than the present age of the Universe  $\left( t_{\mathrm{B}}(r) << t \right)$. Complying with our initial assumption of a homogeneous, early Universe (even in regions from which we do not observe CMB photons), we can thus safely neglect the bang time function and assume that it is zero, $t_{\mathrm{B}}(r) = \mathrm{const} = 0$. This assumption is particularly important for the CMB analysis described in Sect.~\ref{sec:CMB_data}. Without cosmological constant ($\Lambda = 0$), the integral in Eq. \eqref{eq:LTB_age} can be solved parametrically and -- dependent on the sign of the local curvature $E(r)$ -- leads to an elliptic ($E < 0$), parabolic ($E = 0$), or hyberbolic ($E > 0$) evolution. For completeness, we provide these parametric solutions with some more useful expressions in Appendix~\ref{app:LTB}. With a non-zero cosmological constant, no general parametric solution exists. Instead, the elliptic integral in Eq. \eqref{eq:LTB_age} has to be computed numerically, which for certain parameter combinations quickly turns into a difficult numerical problem involving singularities, slow convergence (if at all), and poor precision. However, \citet{2012GReGr..44.2449V} showed that the integral in Eq. \eqref{eq:LTB_age} can be transformed to the so-called Carlson symmetric forms of elliptic integrals. This canonical set of elliptic integrals can be solved using iterative algorithms, which are fast, robust, and quickly converge with machine precision \citep{carlson1995}. We refer to \citet{2012GReGr..44.2449V} for the full derivation, extensive calculations, and many useful details concerning the numerical implementation.

To compare LTB models with observational data, we need to solve for the distance-redshift relation on the past null cone (PNC) of a central observer. Inward radial null geodesics integrated backwards in time are described by the following set of first-order, ordinary differential equations: 

\begin{align}
\label{eq:LTB_dtdr}
\frac{\mathrm{d}t(r)}{\mathrm{d}r} &= - \frac{R' \left[ t(r), r \right]}{\sqrt{1 + 2E(r)}} \; , \\
\label{eq:LTB_dzdr}
\frac{1}{1 + z(r)}\frac{\mathrm{d}z(r)}{\mathrm{d}r} &= \frac{\dot{R}' \left[t(r), r\right]}{\sqrt{1 + 2E(r)}} \; .
\end{align}
These equations can be numerically integrated. After solving for $t(r)$ and $z(r)$ on the PNC, we can numerically invert these relations to arbitrarily transform between $t$, $r$ and $z$. For instance, given a redshift $z$, we can infer the corresponding radius $r(z)$ and compute the time $t[r(z)]$ at which an incoming radial null geodesic was at this position. As can be seen from Eq. \eqref{eq:LTB_metric}, the angular diameter distance in LTB models is simply given by the areal radius function,

\begin{equation}
d_{\mathrm{A}} (z) = R(z) = R \left[ t(z), r(z) \right] \; .
\label{eq:LTB_dang}
\end{equation}
The luminosity distance then follows from the reciprocity theorem \citep{1933PMag...15..761E, 2007GReGr..39.1055E, 2009GReGr..41..581E}:
\begin{equation}
d_\mathrm{L} (z) = (1+z)^2 \, d_{\mathrm{A}}(z) \; . 
\label{eq:LTB_dl}
\end{equation}

These equations briefly summarise the most important quantities that are needed for the following sections. As can be seen, LTB models generally depend on three functions: the bang time function $t_{\mathrm{B}}(r)$, the curvature function $E(r)$, and the mass function $M(r)$. However, the metric and all previous formulae are invariant by diffeomorphism symmetry, including coordinate transformations of the form $r = f(r')$. This gauge freedom can be used to eliminate one function, implying that the physical evolution of LTB models is fully determined by only two free functions. We discussed above that the bang time function can be set to zero by demanding a homogeneous early Universe. This leaves us with only one arbitrary function.  In Sect.~\ref{sec:LTB_algorithm}, we describe in detail how the curvature and mass functions can be derived from a matter density profile at a given time.

\section{Observational data}
\label{sec:obs_data}

This section outlines what observational data we used to constrain LTB models. In this context, two complications have to be considered.

Firstly, many cosmological data sets are routinely reduced under the implicit assumption that our Universe is, on large scales, properly described by an FLRW metric. When fitting alternative cosmological models to these data, special care has to be taken that only model-independent observational constraints are used.

Secondly, linear perturbation theory in LTB models is substantially more complicated than in FLRW models. This is mainly because scalar and tensorial perturbations can couple on inhomogeneous backgrounds. Although there has been great progress in developing a gauge-invariant linear perturbation theory for LTB models, the problem has yet to be fully solved \citep{2008PhRvD..78d3504Z, 2009JCAP...06..025C, 2013arXiv1311.5241F}. To be conservative, we can therefore only use observables that do not depend on the details of structure formation.

Perhaps we should emphasise this limitation of the present work. The current state of the theory only allows us to explore constraints on the global structure of the smooth, unperturbed space-time geometry on large scales. We cannot yet realistically model the evolution of linear perturbations in radially inhomogeneous space-times because we lack proper theoretical tools for predicting\ the statistical properties of the perturbed matter density field,
for example. Observations of the local matter distribution, the large-scale structure (including voids and filaments), cluster number counts or galaxy-galaxy correlation functions can therefore not yet be taken into account for constraining the underlying space-time geometry. We also have to exclude constraints from\ weak-lensing spectra or baryonic acoustic oscillations, for instance, which are widely used to constrain homogeneous and isotropic cosmologies. We are currently working on advancing linear perturbation theory in LTB models, and will include some of the observables listed above in future work.

The lack of linear perturbation theory is not important for Sect.~\ref{sec:results_ltb_zero_lambda}, where we show that only constraints from the local Hubble rate and the CMB are sufficient to rule out LTB models without cosmological constant. For the general probe of the cosmological principle presented in Sect.~\ref{sec:results_lltb}, it would certainly be desirable to have more data. On the other hand, \citet{2014MNRAS.438L...6V} showed that the cosmological observables used in this work are currently the most constraining. 

\subsection{Local Hubble rate}
\label{sec:data_H0}

\citet{2011ApJ...730..119R} used a nearby sample $\left( 0.023 < z < 0.1 \right)$ of Cepheid-calibrated type Ia supernovae to measure the local Hubble rate with a remarkable precision:
\begin{equation}
H_0 = (73.8 \pm 2.4) \, \mathrm{km} \, \mathrm{s}^{-1} \, \mathrm{Mpc}^{-1} \; .
\label{eq:local_hubble_rate}
\end{equation}
Up to now, this is probably the most accurate determination of the local expansion rate of the Universe. Although \citet{2014MNRAS.440.1138E} recently reanalysed the data from \citet{2011ApJ...730..119R} and proposed a corrected, slightly lower value of $H_0 = (72.5 \pm 2.5) \, \mathrm{km} \, \mathrm{s}^{-1} \, \mathrm{Mpc}^{-1}$; we used the original measurement to constrain our models because the proposed corrections are not significant for our main conclusions.

One essential property of LTB models is their radially dependent expansion rate. More precisely, we can define a longitudinal Hubble rate, $H_{\mathrm{L}}(t,r) = \dot{R}'(t,r)/R'(t,r)$, which describes the expansion rate along the radial direction, and a transversal Hubble rate, $H_{\mathrm{T}}(t,r) = \dot{R}(t,r)/R(t,r)$, which describes the expansion rate of the individual spherical shells. It is therefore not a priori clear how LTB models should be compared with the above-mentioned measurement. 

To mimic the procedure of \citet{2011ApJ...730..119R}, we used the following approach: Independent of the cosmological model, the (local) luminosity distance can be considered an analytic function of redshift $z$ and hence expanded in a Taylor series,

\begin{align}
d_{\mathrm{L}}(z) = \frac{c}{H_0} & \left[ z + \frac{1}{2} \, (1-q_0) \, z^2 - \frac{1}{6} \, (1 - q_0 - 3 q_0^2 + j_0) \, z^3  \right] \notag \\
 & + \mathcal{O}(z^4) \qquad (z \ll 1) \; ,
\label{eq:dl_taylor_series}
\end{align} 
where $q_0$ and $j_0$ are the deceleration parameter and the jerk, respectively. Like \citet{2011ApJ...730..119R}, we fixed these two parameters to $q_0 = -0.55$ and $j_0 = 1$. To compute the \textit{effective} local Hubble rate of a specific LTB model, we first tabulated the luminosity distance $d_{\mathrm{L}}(z_i)$ (cf. Eq. \eqref{eq:LTB_dl}) at $N$ equidistantly spaced steps in the considered redshift range, $0.023 < z_1 < ... < z_N < 0.1$, and then calculated the best-fitting (least-squares) Hubble rate $H_{\mathrm{LS}}$ through these data points using Eq. \eqref{eq:dl_taylor_series}. The deviation from the observed value is then quantified with a simple chi-square,
\begin{equation}
\chi_{H_0}^2 = \dfrac{ \left( H_{\mathrm{LS}} - H_0 \right)^2}{\sigma_{H_0}^2} \qquad \left( \sigma_{H_0} = 2.4 \, \mathrm{km} \, \mathrm{s}^{-1} \, \mathrm{Mpc}^{-1} \right) \; .
\end{equation}
Using this approach, we defined an averaged, effective local Hubble rate for LTB models that closely mimics the one measured with observed type Ia supernovae.

\subsection{Supernovae}
\label{sec:data_SNe}

To constrain the shape of the luminosity distance at even higher redshifts, we used the Union2.1 compilation released by the Supernova Cosmology Project \citep{2012ApJ...746...85S}. This catalogue contains 580 uniformly analysed type Ia supernovae and extends out to redshift $z \sim 1.5$. Currently, the Union2.1 compilation is the largest, and most recent, publicly available sample of standardized type Ia supernovae.

Although the shape of type Ia supernovae light-curves is empirically well understood, their absolute magnitude is essentially unknown and needs to be calibrated. Samples like the Union2.1 compilation are therefore reduced by fixing the Hubble rate to an arbitrary value. It is important to remove this artificial constraint from the data when fitting cosmological models. This can either be achieved by analytically marginalizing over the assumed Hubble rate -- or, equivalently, the absolute magnitude -- (see \citet{2002MNRAS.335.1193B} or Appendix C.2 of \citet{2010JCAP...11..030B}), or by using the elegant weight matrix formalism described in the appendix of \citet{2010ApJ...716..712A}. The weight matrix is constructed from the full covariance matrix with systematics (e.g. host mass correction) and incorporates the marginalization over various nuisance parameters. In particular, the marginalization over the Hubble rate is included. We use the weight matrix formalism to perform likelihood estimates in the following sections.

Finally, we note that several authors argued that supernova samples reduced with the SALT-II light-curve fitter from \citet{2007A&A...466...11G} are systematically biased towards the standard cosmological model and tend to disfavour alternative cosmologies \citep{2009ApJ...700.1097H, 2009ApJS..185...32K, 2011MNRAS.413..367S}. The Union2.1 compilation was reduced with the SALT-II fitter, and hence this potential penalty would also affect the goodness-of-fit of LTB models. However, we can safely neglect this problem for mainly two reasons. Firstly, we did not conduct a detailed, statistical comparison (e.g. Bayesian model comparison) between the standard cosmological model and LTB models here, and therefore the small, systematic corrections are irrelevant for our main conclusions. Secondly, and more importantly, we demonstrate that the supernova data do not impose tight constraints on LTB models anyway. Instead, supernovae can easily be fitted with a variety of different density profiles and certainly do not cause the tension between observations and LTB models that we focus on.

\subsection{Cosmic microwave background}
\label{sec:CMB_data}

The standard approach for analysing CMB spectra is inherently based on the assumption of a spatially isotropic and homogeneous Universe \citep{2013ApJS..208...20B, 2013arXiv1303.5076P}. Primary anisotropies are calculated by numerically solving the Boltzmann equations on an FLRW background \citep{2000ApJ...538..473L, 2011arXiv1104.2932L}. Secondary anisotropies are caused by different forms of interactions between cosmic structures and the CMB photons after the time of decoupling, such as the late integrated Sachs-Wolfe (ISW) effect or rescattering during reionisation. As such, all these processes depend on the details of structure formation and are commonly modelled using perturbation theory in FLRW models. It is obvious that our ansatz for the metric in Sect.~\ref{sec:LTB_metric} strongly violates these usual assumptions. How then can we self-consistently use CMB spectra to constrain LTB models?

We recall that we wish to retain the inflationary paradigm and assume that the early Universe is highly homogeneous. This means that, even in our approach, the early Universe can effectively be described by a FLRW metric. Thus, the modelling of primary CMB anisotropies does not differ from the standard approach at least until the time of decoupling. This was also the motivation for excluding variations of the bang time function $t_{\mathrm{B}}(r)$ from Eq. \eqref{eq:LTB_age}. If we were to drop the assumption of a homogeneous early Universe, we would have to develop a general relativistic formalism for calculating CMB anisotropies on inhomogeneous backgrounds. Clearly, this would go far beyond the scope of this work.

Properly treating secondary CMB anisotropies in LTB models is, however, more complicated because linear perturbation theory is still being developed and cannot yet be used to calculate these secondary effects. We can circumvent this problem by following the elegant work of \citet{2010JCAP...08..023V}, which describes a method for analysing the CMB in a manner that is as independent as possible of late-time cosmology. To this end, the authors begin with identifying the three dominant imprints that the late cosmological model leaves on the observed CMB spectrum. Firstly, CMB photons are lensed as they traverse non-linear structures. The late ISW effect dominates the CMB spectrum on large scales ($l \lesssim 40$). Consequently, low multipoles strongly depend on the detailed properties of the late cosmological model. Secondly, the overall amplitude of the CMB spectrum at $l \gg 40$ is reduced by scattering processes during the epoch of reionisation. This suppression is usually parametrised by the factor $e^{-2 \tau}$, where $\tau$ is the reionisation optical depth. Thirdly, the angular diameter distance to the last scattering surface (LSS) determines the angular scales of the acoustic peaks. This is a simple projection effect. Variations of the angular diameter distance shift the CMB spectrum in multipole space. 

If large scales ($l \lesssim 40$) are excluded from the CMB analysis, the dominant perturbations of the primordial CMB spectrum due to the late cosmological model can therefore be parametrised by a rescaling factor $\alpha$ of the global amplitude and a shift parameter $\beta$. Rephrased more mathematically, the expansion coefficients of the CMB power-spectrum in multipole space transform as ${C_l  \rightarrow \alpha C_{\beta l}}$. 

This simple approximation is already astonishingly accurate for large parts of the CMB spectrum. But the crucial point is that this insight allowed \citet{2010JCAP...08..023V} to encode unknown secondary effects in carefully chosen nuisance parameters (e.g. the global amplitude of the CMB spectrum). These additional parameters can be built into standard parameter estimation codes used for analysing CMB spectra. To derive minimal constraints that do not depend on the detailed properties of the late cosmological model, one simply has to marginalise over these newly introduced nuisance parameters. 

Except for excluding high multipoles ($l > 800$) from the analysis, \citet{2010JCAP...08..023V} neglected the impact of gravitational lensing on the CMB spectrum. This was well justified given the accuracy of CMB experiments at the time of publishing. Motivated by the improved accuracy of modern CMB experiments, however, \citet{2013JCAP...02..001A} extended the original method of \citet{2010JCAP...08..023V} by introducing a new technique for additionally marginalising over the CMB lensing contamination at all multipole orders. Initially, we used this advanced technique in combination with the publicly available parameter estimation code \texttt{Monte Python} \citep{2013ascl.soft07002A} to derive model-independent constraints from the latest Planck data \citep{2013arXiv1303.5076P}. While we prepared this manuscript, \citet{2013arXiv1312.5696A} published another slight refinement of their own method and also applied it to the Planck 2013 data. We  therefore used the results published in \citet{2013arXiv1312.5696A} to estimate likelihoods in the following sections. The relevant model-independent constraints including one-sigma errors are summarised in Table~\ref{tab:CMB_constraints}.

\begin{table}
\caption[Model-independent constraints from the Planck 2013 data.]{Model-independent constraints from the Planck 2013 data published by \citet{2013arXiv1312.5696A}. $\omega_{\mathrm{b}} = \Omega_{\mathrm{b}} h^2$ and $\omega_{\mathrm{c}} = \Omega_{\mathrm{c}} h^2$ denote the physical baryon and cold dark matter densities, respectively. $d_{\mathrm{A}}(z_*)$ is the angular diameter distance to the surface of last scattering.}
\label{tab:CMB_constraints}
\centering
\begin{tabular}{c|c|c}
$ 100 \, w_{\mathrm{b}}$ & $w_{\mathrm{c}}$ & $d_{\mathrm{A}}(z_*) \, [\mathrm{Mpc}]$ \\
\hline \Tstrut
$2.243 \pm 0.040$ & $0.1165 \pm 0.0037$ & $12.80 \pm 0.068$
\end{tabular}
\end{table}

Finally, we need to explain how LTB models can actually be fitted to these constraints.  We begin by noting that the redshift of decoupling $z_*$ generally is a function of the physical matter densities $w_{\mathrm{b}}$ and $w_{\mathrm{c}}$. Assuming standard radiation content, however, this dependence is only weak \citep{1996ApJ...471..542H}. For simplicity, we can thus fix the decoupling redshift to $z_* = 1090$. Following previous works \citep[see e.g.][for detailed descriptions]{2008PhRvL.101y1303Z, 2010JCAP...11..030B, 2011PhRvD..83j3515M}, we used an effective FLRW observer approach for computing the angular diameter distance $d_{\mathrm{A}}(z_*)$ to the LSS: Radial null geodesics are numerically integrated in LTB models only up to an embedding redshift $z_{\mathrm{b}}$. This redshift has to be chosen such that the LTB models are already sufficiently homogeneous (i.e. the gravitational shear $\sigma^2 = \frac{2}{3} \left( H_{\mathrm{L}} - H_{\mathrm{T}} \right)$ vanishes) and radiation is still negligible. Typically, $z_{\mathrm{b}} \approx 150$ fulfils these criteria. 

After reaching this embedding redshift, we continued to compute the distance-redshift relation in an effective FLRW background model up to the decoupling redshift $z_* = 1090$. The effective FLRW model was chosen such that a fictitious observer in this background model would observe the same CMB spectrum. This approach is beneficial because (1) solving the distance-redshift relation in FLRW models is computationally substantially cheaper, and (2) radiation can be included. Lastly, the appropriately scaled matter densities and the calculated angular diameter distance can be compared with the model-independent constraints from Table~\ref{tab:CMB_constraints} using a simple chi-square.

\subsection{Kinetic Sunyaev-Zel'dovich}
\label{sec:ksz_data}

\begin{figure}
	\centering
	\includegraphics[width=0.45\textwidth]{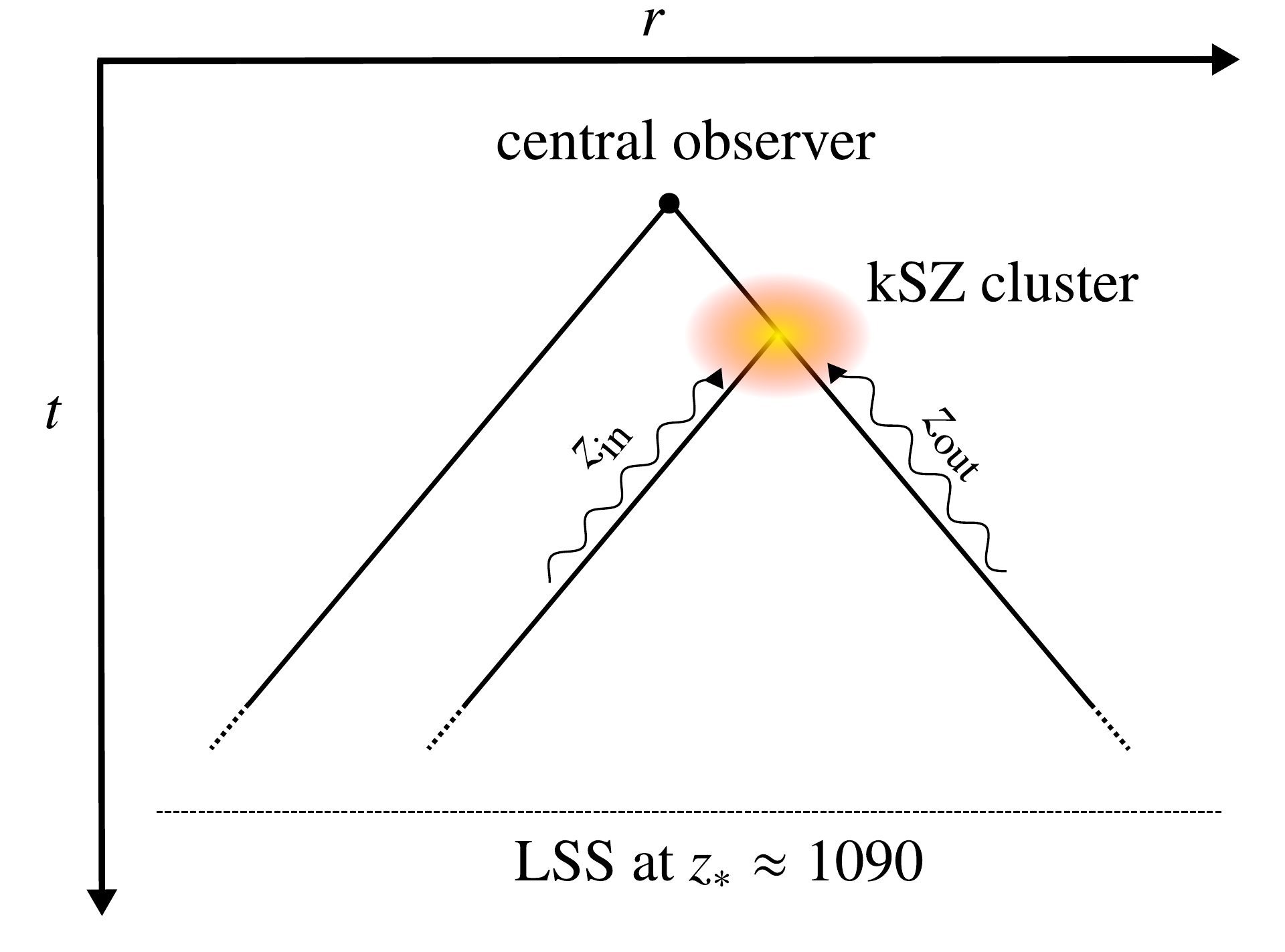}
	\caption[Kinetic Sunyaev-Zel'dovich effect in LTB models.]{Apparent kinetic Sunyaev-Zel'dovich (kSZ) effect of off-centre galaxy clusters in radially inhomogeneous LTB models. The redshift $z_{\mathrm{in}}$ of CMB photons that propagated through the centre of the inhomogeneity generally differs from the redshift $z_{\mathrm{out}}$ of CMB photons that arrived from outside.}
	\label{fig:LTB_ksz_effect}
\end{figure}

Cosmic microwave background photons traversing galaxy clusters interact with the hot intra-cluster gas through inverse Compton scattering. These interactions cause the Sunyaev-Zel'dovich effect, which manifests itself by characteristic distortions of the energy spectrum of the rescattered photons. The dominant contribution in galaxy clusters is due to the thermal Sunyaev-Zel'dovich effect \citep{1970Ap&SS...7....3S, 1972CoASP...4..173S}: thermal energy is transferred from the hot intra-cluster gas to the CMB photons, causing a redistribution of photons from lower to higher energy states. A second-order contribution is due to the kinetic Sunyaev-Zel'dovich (kSZ) effect \citep{1980MNRAS.190..413S}: galaxy clusters moving with a non-zero peculiar velocity with respect to the rest frame of the CMB observe an anisotropic CMB spectrum with non-zero dipole moment. This dipole induces a characteristic shift in the spectrum of the reflected CMB photons, which is similar to the relativistic Doppler effect. In principal, the kSZ effect of galaxy clusters can therefore be used to measure their peculiar velocities relative to the CMB.

As was first shown by \citet{2008JCAP...09..016G}, the kSZ effect can be used as a powerful probe of radially inhomogeneous LTB models. To understand this, we recall that LTB models are isotropic only about the symmetry centre at the coordinate origin. Off-centre observers therefore generally see an anisotropic CMB spectrum \citep[see][for detailed calculations]{2006PhRvD..74j3520A}. To first order, the anisotropy can be approximated as a pure dipole, which because of the symmetry of the problem is aligned along the radial direction. This effect is depicted in Fig.~\ref{fig:LTB_ksz_effect}, which shows an off-centre galaxy cluster that rescatters CMB photons that arrive from the LSS. 

The extreme redshifts (as seen from the galaxy cluster) are observed along the radial direction, for CMB photons with redshift $z_{\mathrm{in}}$ that propagated through the centre of the inhomogeneity and, in the opposite direction, for CMB photons with redshift $z_{\mathrm{out}}$ arriving from outside. In the case of a giant void scenario, for instance, photons arriving from inside the void travelled the longest distance through an underdense region. Consequently, they also spent the longest time in a space-time region with a higher expansion rate, so that their redshift $z_{\mathrm{in}}$ reaches the highest possible value. Vice versa, CMB photons arriving from outside the void exhibit the lowest redshift $z_{\mathrm{out}}$. 

Note that the kSZ effect described here is only an apparent kSZ effect; it is not caused by real peculiar motions of the galaxy clusters. Instead, the effect only appears because the space-time geometry around off-centre observers is anisotropic. Note also that the kSZ effect has a distinguished feature in comparison with most other cosmological probes. The galaxy clusters act as mirrors for CMB photons, reflecting radiation from all spatial directions. By analysing the spectrum of the reflected light, we can therefore extract information about space-time regions that would otherwise be inaccessible. In this sense, by measuring the difference between the redshift $z_{\mathrm{in}}$ and $z_{\mathrm{out}}$, the kSZ effect allows us to (indirectly) look inside our past-null cone (cf. Fig.~\ref{fig:LTB_ksz_effect}).

As already mentioned, to first order, the anisotropy observed by off-centre galaxy clusters can be approximated as a pure dipole. This approximation is sufficiently accurate as long as the effective size of the LTB inhomogeneity on the sky, as observed by the galaxy cluster, is larger than $\sim 2 \pi$ \citep{2006PhRvD..74j3520A}. The amplitude of the dipole is then given by
\begin{equation}
\label{eq:ltb_ksz_dipole}
\beta = \frac{v}{c} = \frac{\Delta T}{T} = \frac{z_{\mathrm{in}} - z_{\mathrm{out}}}{2 + z_{\mathrm{in}} + z_{\mathrm{out}}} \; .
\end{equation}
Measurements of the kSZ effect of individual galaxy clusters are extremely difficult and suffer from low signal-to-noise ratios. Current data exhibit very large errors and therefore still need
to be considered premature. However, even though the uncertainties are huge, \citet{2008JCAP...09..016G} and \citet{2012PhRvD..85b4002B} showed that the currently available measurements already place tight constraints on the depth and radial size of Gpc LTB voids. In this work, we use the kSZ data of the nine galaxy clusters compiled by \citet{2008JCAP...09..016G}, assuming a conservative scatter of $\sigma_{\mathrm{pv}} = 1600 \, \mathrm{km} \, \mathrm{s}^{-1}$ and zero systematic shift, $v_{\mathrm{sys}} = 0$; see \citet{2008JCAP...09..016G} and references therein for a detailed discussion of the data, sources of errors, and the modelling of the likelihood. 

To compute the expected kSZ effect for a given galaxy cluster, we first determined the cluster coordinates $(t_{\mathrm{cl}}, r_{\mathrm{cl}}, z_{\mathrm{cl}})$ on the past-null cone of the central observer. Starting from this position, we then solved for ingoing and outgoing radial null geodesics by numerically integrating Eqs. \eqref{eq:LTB_dtdr} and \eqref{eq:LTB_dzdr} up to the LSS. This procedure yields the two redshifts  $z_{\mathrm{in}}$ and $z_{\mathrm{out}}$, which quantify the CMB dipole as seen by the off-centre galaxy cluster.

\section{Monte Carlo approach for constraining the local density profile}
\label{sec:LTB_statistical_approach}

Most studies published so far have assumed certain functional forms for either the mass profile or the curvature profile of LTB models that may be considered characteristic for voids \citep[see e.g.][for typical profiles]{2008JCAP...04..003G, 2009JCAP...02..020B, 2010JCAP...11..030B}. Of course, these empirical functional forms impose artificial constraints on the models when performing maximum likelihood estimates.

Our approach here is different because we parametrise the matter density profile of LTB models as flexibly as possible, imposing few a priori constraints on its detailed shape. To derive statistical constraints, we then use a parameter estimation code to systematically vary the density profile. We wish to determine the shape of the favoured profile, and also how tight the constraints on the detailed shape are. We also investigate whether highly flexible profiles allow us to mitigate the reported tension between measurements of the local Hubble rate and the CMB data \citep{2010JCAP...11..030B, 2011PhRvD..83j3515M, 2012PhRvD..85b4002B, 2012JCAP...10..009Z}. To this end, we proceed by first discussing our choices for the parametrisation of the local density profile, continue with the algorithm that allows us to derive LTB models from these profiles, and conclude this section by explaining some details of the parameter estimation technique.

\subsection{Flexible models for the local density profile}

\begin{figure}
	\centering
	\includegraphics[width=0.45\textwidth]{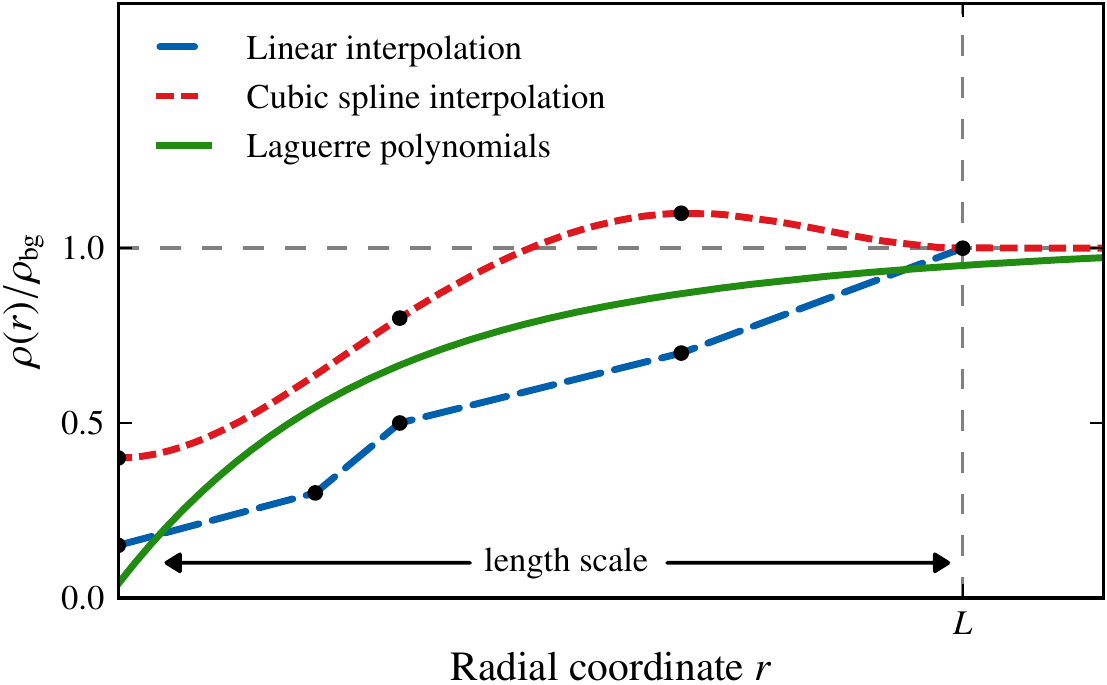}
	\caption[Visualisation of different models for LTB density profiles.]{Illustration of the three different parametrisations for the local matter density profile around the observer. The black dots indicate the nodes of the interpolation schemes. Note that the representation by Laguerre polynomials does not have a fixed radial size, but only asymptotes to the homogeneous background density.}
	\label{fig:density_models}
\end{figure}

Broadly speaking, we have two options to parametrise the local density profile in the most flexible way. We can represent the density profile by a general interpolation scheme, or alternatively decompose the void profile function into a series of appropriately chosen, orthonormal basis functions of the radial coordinate. After some initial testing, we decided to use the following parametrisations:

\begin{enumerate}

\item \textbf{Linear interpolation}: The first free model parameter is the radial size $L$ of the inhomogeneity. For radii $r < L$, we use linear interpolation, specified by pairs of radial coordinates and corresponding matter densities: [$(r_0 = 0, \rho_0)$, $(r_1, \rho_1)$, ..., $(r_n, \rho_n)$], where $r_n < L$ and all densities $\rho_i$ are strictly positive. For radii $r \geq L$, the density profile is constant and equals the matter density $\rho_{\mathrm{bg}}$ of the homogeneous FLRW background. $M'(r)$ (cf. Eq. \eqref{eq:LTB_EFE2}), and thereby also radial derivatives of the areal radius function $R(t,r)$ (cf. Eq. \eqref{eq:LTB_R_prime}) are not smooth if the density profile is linearly interpolated. All functions required for this work remain continuous, however, which is sufficient for the computations. \\ 

\item \textbf{Cubic spline interpolation}: This model has the same free parameters as the linear interpolation model: a radial size $L$,  pairs of radial coordinates, and corresponding matter densities $\left[ r_i, \rho_i = \rho(r_i) \right]$. Cubic splines are used to interpolate between these nodes. We chose the boundary conditions $\rho(r \geq L) = \rho_{\mathrm{bg}}$ and $\rho'(r = 0) = \rho'(L) = 0$ to enforce a smooth void profile at the origin and guarantee a smooth embedding into the homogeneous FLRW background. This is essentially equivalent to the spline model used by \citet{2008PhRvL.101y1303Z} and \citet{2011PhRvD..83j3515M}. \\ 

\item \textbf{Laguerre polynomials}: Again, the first free model parameter is a radial length scale $L$. We introduce the scaled radial coordinate $x = r/L$ and define the matter density profile as
\begin{equation}
\rho(t_0, r) = \rho_{\mathrm{bg}} \left[1 + \exp(-x) \sum_i a_i \mathrm{L}_i(x)  \right] \; ,
\end{equation}
where $\mathrm{L}_i$ denote the Laguerre polynomials, which are an orthogonal basis with respect to the inner product $\langle \mathrm{L}_i, \mathrm{L}_j \rangle = \int_0^{\infty} \mathrm{L}_i(x) \, \mathrm{L}_j(x) \, \exp(-x) \, \mathrm{d}x$. In practice, we used Laguerre polynomials up to fifth order, because more degrees of freedom were not constrained by the data. In contrast
to the previous interpolation schemes, the Laguerre models do not have a fixed size. Instead, their density profiles only asymptote (characteristic length scale $L$) to the homogeneous background density.

\end{enumerate}

The different approaches are visualised in Fig.~\ref{fig:density_models}. We conducted detailed tests as explained in Sect.~\ref{sec:results_ltb_zero_lambda} with each of the three parametrisations, finding qualitatively the same results. Because the meaning of the free model parameters of the interpolation schemes is quite instructive while on the other hand the geometrical interpretation of expansion coefficients of Laguerre polynomials is not straightforward, we focus our discussion in the subsequent sections on the linear and cubic spline interpolations.

\subsection{Algorithm}
\label{sec:LTB_algorithm}

In our approach, LTB models are generally determined by the dimensionless Hubble parameter $h = H_0/(100 \, \mathrm{km} \, \mathrm{s}^{-1} \, \mathrm{Mpc}^{-1})$ and the physical densities of baryonic and cold dark matter, $\omega_{\mathrm{b}} = \Omega_{\mathrm{b}} h^2$ and $\omega_{\mathrm{c}} = \Omega_{\mathrm{c}} h^2$, of the homogeneous FLRW background, as well as a list of model parameters $(a_1, ..., a_n)$ that parametrise the local matter density profile $\rho(t_0, r)$ at the present time $t_0$. If not stated otherwise, these are the base parameters that are later sampled by the Monte Carlo method that was used for fitting LTB models to observational data. Given these base parameters, our algorithm for computing observables in the LTB metric can be outlined as follows:

\begin{enumerate}

\item \textbf{Age of the Universe}: As emphasised before, we demand a homogeneous early Universe and thus set the bang time function $t_{\mathrm{B}}(r)$ to zero. Therefore, the LTB patch and the homogeneous background have the same global age $t_0$, which we compute using the standard FLRW relation,

\begin{equation}
t_0 = \frac{1}{H_0} \int_0^1 \frac{\sqrt{a}}{\sqrt{\Omega_m+\Omega_k a}} \, \mathrm{d}a \; ,
\end{equation}
where the curvature parameter is given by $\Omega_{\mathrm{k}} = 1 - \Omega_{\mathrm{m}}$, that is, we explicitly consider curved FLRW backgrounds.

\item \textbf{Gauge freedom}: All formulae given in Sect.~\ref{sec:LTB_metric} are invariant under coordinate transformations of the form $\tilde{r} = g(r)$. We exploit this gauge freedom to scale the radial coordinate $r$ such that it equals the areal radius function at the present time: $R(t_0, r) = r$.

\item \textbf{Mass profile}: Given the parameters $(a_1, ..., a_n)$, we construct the matter density profile at the present time $\rho(t_0, r)$. We reject combinations of parameters that induce negative matter densities. This is important for the cubic spline interpolation and the Laguerre polynomials. In our gauge, the effective gravitational mass $M(r)$ then directly follows from integrating Eq. \eqref{eq:LTB_EFE2},

\begin{equation}
M(r) = 4 \pi G \int \limits_0^r \rho(t_0, \tilde{r}) \, \mathrm{d}\tilde{r} \; . 
\end{equation}

\item \textbf{Curvature profile}: The curvature function $E(r)$ is implicitly defined by Eq. \eqref{eq:LTB_age} and can only be computed numerically. We use the TOMS 748 root-bracketing algorithm from \citet{Alefeld:1995:AEZ:210089.210111} to determine $E(r)$ as a function of $t_0$, $r$ and $M(r)$, see \citet{2002PhRvD..65b3501K} for details on choosing the initial bracket. We reject mass models that require $E(r) \leq -1/2$ because this would cause singular line-elements (cf. Eq. \eqref{eq:LTB_metric}). The radial derivative $E'(r)$ is numerically computed using a standard fourth-order finite differencing scheme.

\item \textbf{Distance-redshift relation}: We use a fifth order Dormand-Prince method -- which is essentially a Runge-Kutta scheme with error control and adaptive step size \citep{odeint_paper} -- to numerically integrate the ordinary differential equations describing radial null geodesics on the PNC of a central observer (cf. Eqs. \eqref{eq:LTB_dtdr} and \eqref{eq:LTB_dzdr}). The resulting relations $t(r)$ and $z(r)$ are then interpolated and numerically inverted using smooth Akima splines \citep{Akima:1970:NMI:321607.321609}. This allows us to arbitrarily transform between $t$, $r$ and $z$ and, in particular, to compute angular diameter and luminosity distances (see Sect.~\ref{sec:LTB_metric}). We discard LTB models that exhibit shell-crossings or multivalued redshifts on the PNC \citep{1985ApJ...290..381H}.

\end{enumerate}

This algorithm allowed us to compute all observable quantities that are required for performing likelihood estimates of LTB models.

\subsection{Efficient statistical sampling}

We are confronted with a typical parameter estimation problem. Given observational data, we need to explore the posterior distribution in a high-dimensional parameter space to estimate the most likely values of the free model parameters $(h, \omega_{\mathrm{b}}, \omega_{\mathrm{c}}, a_1, ..., a_n)$. Monte Carlo methods are the standard approach for solving this kind of problem, and most commonly, variants of the simple Metropolis-Hastings algorithm are used \citep{hastings1970monte}.

The Metropolis-Hastings algorithm has an important drawback,
however: it requires a fine-tuned proposal distribution to efficiently sample the posterior. If the proposal distribution is thought of as a multivariate Gaussian, this means that each entry of the covariance matrix needs to be tuned. In our case, the matter densities at different radial coordinates can be correlated and hence the covariance matrix is non-diagonal. Consequently, we would have to hand-tune $N(N+1)/2$ unknown parameters (where $N$ is the dimension of the parameter space). This is an extremely time-consuming task, in particular because the fine-tuned parameters strongly depend on the precise parametrisation (e.g. number and position of interpolation nodes) of the matter density profile.

After some testing, we decided to use an alternative Monte Carlo sampler: the so-called stretch-move technique, which was first introduced by \citet{goodman2010ensemble}. This technique is affine-invariant, meaning that it performs equally well under all linear transformations of the parameter space. In particular, it is insensitive to covariances between parameters and therefore requires no fine-tuning. \citet{goodman2010ensemble} demonstrated the excellent performance of their algorithm (as measured by the auto-correlation time) for several pathological posterior distributions.

In addition, the stretch-move sampler simultaneously explores the parameter space with a whole ensemble of Monte Carlo walkers. The time evolution of this ensemble can easily be parallelised, which greatly reduces the required computing time (wall-clock time) on multi-core machines or large computing clusters \citep{2013A&C.....2...27A}. \citet{2013PASP..125..306F} provided an excellent discussion of the stretch-move technique and described a parallelised implementation in detail. Following this, we implemented the ensemble sampler in \texttt{C++} and parallelised it with the message-passing interface (MPI) system. This enabled us to perform likelihood estimates of many different LTB models within a significantly shorter computing time. 

All parameter estimations in the remainder of this work were performed with the stretch-move technique, typically using hundreds of walkers. The length of the burn-in period was estimated by means of the exponential auto-correlation time $\tau_{\mathrm{exp}}$. Convergence of the samples was ensured by letting the random walks proceed for multiple integrated auto-correlation times $\tau_{\mathrm{int}}$ after removing all samples drawn during the initial burn-in period. This procedure was advocated by \citet{2013A&C.....2...27A} and \citet{2014MNRAS.437.3918A}, who provided a detailed discussion of convergence diagnostics with the auto-correlation times described above.

\section{Comparing LTB models without cosmological constant with observational data}
\label{sec:results_ltb_zero_lambda}

We now compare LTB models with consecutively different combinations of observational data. This stepwise analysis allows us to carefully explain why LTB models with a constant bang time function and zero cosmological constant are inconsistent with current data. The following results are representative. For each scenario discussed, we fitted numerous different LTB models, varying the radial size of voids, changing the numbers and positions of interpolation nodes, or considering different orders of the expansion in terms of the Laguerre polynomials. Our findings with the different approaches agreed qualitatively well, therefore we only discuss simple parametrisations that already show all important characteristics. 

\subsection{Constraints: $H_0$ + supernovae}
\label{sec:SNe_Hubble}

To begin with, we only analysed the constraints imposed by the local Hubble rate and supernovae. We therefore considered LTB models of fixed radial size $L = 3 \, \mathrm{Gpc}$ and parametrised the matter density profiles with linear and cubic spline interpolation schemes with three equidistant nodes at radii $r_1 = 0$, $r_2 = 1  \, \mathrm{Gpc}$, $r_3 = 2  \, \mathrm{Gpc}$. It is convenient to express the matter densities at these nodes with respect to the matter density $\rho_{\mathrm{bg}}$ of the FLRW background model, viz $a_i = \rho(r_i)/\rho_{\mathrm{bg}}$. The priors for the parameters $a_i$ are only bounded from below ($a_i \geq 0$), so that the stretch-move walkers can essentially freely explore the physically relevant parameter space.

Anticipating the following results, we asymptotically embedded the LTB models in Einstein-de-Sitter (EdS) backgrounds for this particular test, meaning that we explicitly set the spatial curvature of the homogeneous background to zero ($\Omega_{\mathrm{k}} = 0$). We did this to demonstrate that good fits to the data can be achieved even with asymptotically flat backgrounds. As can easily be verified, the models considered are fully determined by the physical matter density $\omega_{\mathrm{m}}$ of the EdS background and the profile parameters $(a_1, a_2, a_3)$.

\begin{figure}
	\centering
	\includegraphics[width=0.45\textwidth]{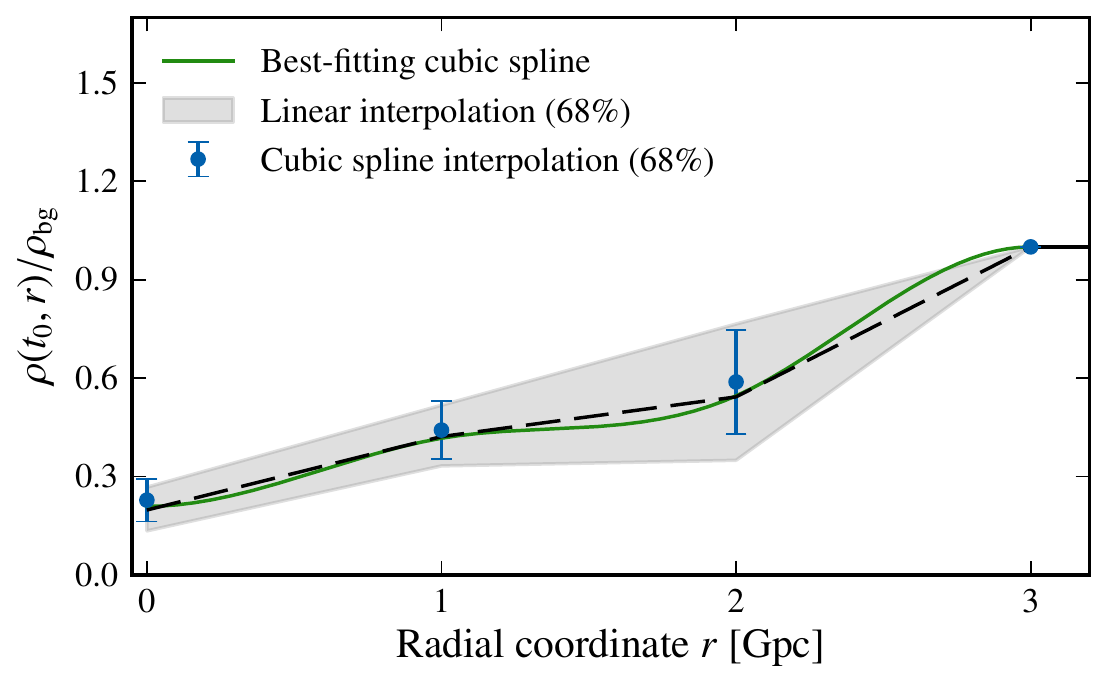}
	\caption[Best-fitting matter density profiles given measurements of the local Hubble rate and supernovae.]{Statistical constraints on the shape of the local matter density profile at the present time. LTB models with fixed radial size $L = 3 \; \mathrm{Gpc}$ and three equidistant interpolation nodes were constrained by measurements of the local Hubble rate and supernovae. The blue error bars and the grey-shaded band indicate the 68\% confidence intervals for the cubic spline and linear interpolation schemes, respectively. The black dashed line connects the means of the best-fitting nodes for the linear interpolation scheme. The green solid curve shows the best-fitting cubic spline density profile.}
	\label{fig:SNe_Hubble_bf_profiles}
\end{figure}

Figure~\ref{fig:SNe_Hubble_bf_profiles} illustrates the preferred shape of the local matter density profile given the local Hubble rate and supernova data. More precisely, we show the 68\% confidence intervals for the best-fitting linear and cubic spline interpolation models. Note that the constraints for the two different approaches are almost identical. As expected, the data clearly favour large and deep voids whose density profiles gradually decrease towards the origin. This shape is required to account for the apparent acceleration suggested by the supernova data. Typically, the density contrast at the origin is  $\delta \rho(t_0, r = 0)/\rho_{\mathrm{bg}} \approx -0.75$. It can also be seen that the constraints on the matter density profile weaken at larger radii. This is naturally caused by the quality of the data. Small radii correspond to low redshifts. In this range, the profiles are simultaneously constrained by the measurement of the local Hubble rate ($z < 0.1$) and also by many supernovae with comparably small error bars. At higher redshifts, the amount of observed supernovae decreases, and at the same time, their errors increase (mainly because the supernovae become fainter). Therefore, the matter density profile can substantially fluctuate at larger radii without being penalised too strongly. This freedom also demonstrates that the LTB models considered can easily fit the data without a cosmological constant and zero background curvature. The best-fitting cubic spline model has an excellent log-likelihood value of $\log \left( \mathcal{L} \right) = -272.27$ and fits the data just as well as the best-fitting $\Lambda$CDM model with $\log \left( \mathcal{L} \right) = -272.56$.

\begin{figure*}
	\centering
	\includegraphics[width=0.95\textwidth]{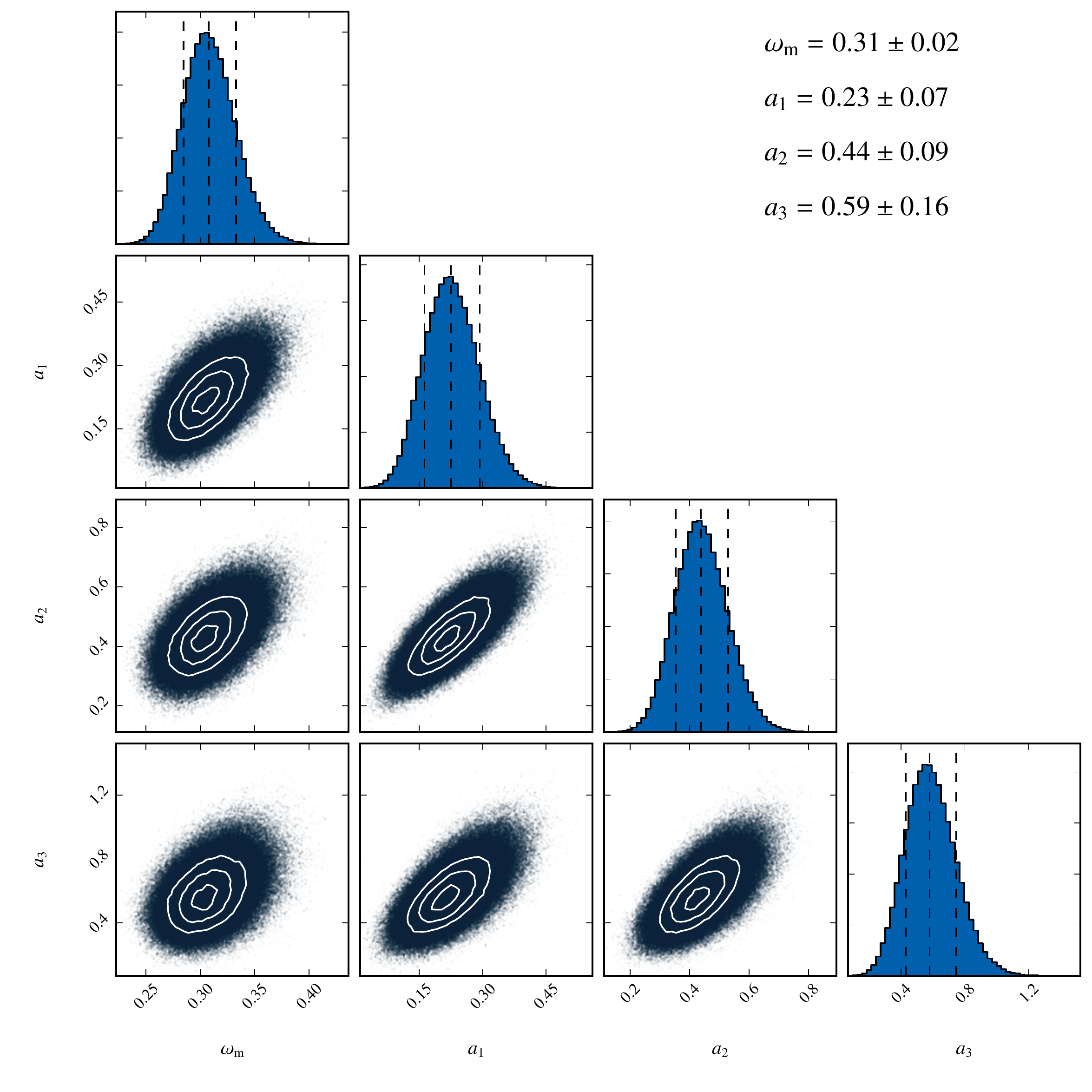}
	\caption[Constraints on LTB models from the local Hubble rate and supernovae.]{Statistical constraints on the shape of the local matter density profile imposed by measurements of the local Hubble rate and supernovae. $\omega_{\mathrm{m}}$ denotes the physical matter density of the EdS background model. The local density profile was parametrised by a cubic spline interpolation with three equidistant nodes at radii $r_1 = 0$, $r_2 = 1  \, \mathrm{Gpc}$, $r_3 = 2  \, \mathrm{Gpc}$ and a fixed radial size of $L = 3  \, \mathrm{Gpc}$. The matter density at these nodes was normalised with respect to the background density today, viz $a_i = \rho(t_0, r_i)/\rho_{\mathrm{bg}}(t_0)$.}
	\label{fig:SNe_Hubble_triangle}
\end{figure*}

Figure~\ref{fig:SNe_Hubble_triangle} reveals more details of the Markov Chain Monte Carlo (MCMC) simulation specifically for the cubic spline model. While the marginalised posterior distributions, the means, and the standard deviations of the density profile parameters $a_i$ confirm the above discussion, the almost perfect linear correlations between the parameters enforce the idea of a gradually decreasing density profile towards the symmetry centre: If $a_i$ is increased, also $a_{i+1}$ has to grow to generate the gradient of the expansion rate required by the data. The physical matter density $\omega_{\mathrm{m}}$ determines the expansion rate of the background EdS model through $h = \sqrt{\omega_{\mathrm{m}}}$. This is why $\omega_{\mathrm{m}}$ and the profile parameters $a_i$ are correlated: If $\omega_{\mathrm{m}}$ decreases, so does the expansion rate of the background model, which means that the voids need to become even deeper to maintain the required high local Hubble rate at the origin. 

\begin{figure}
	\centering
	\includegraphics[width=0.45\textwidth]{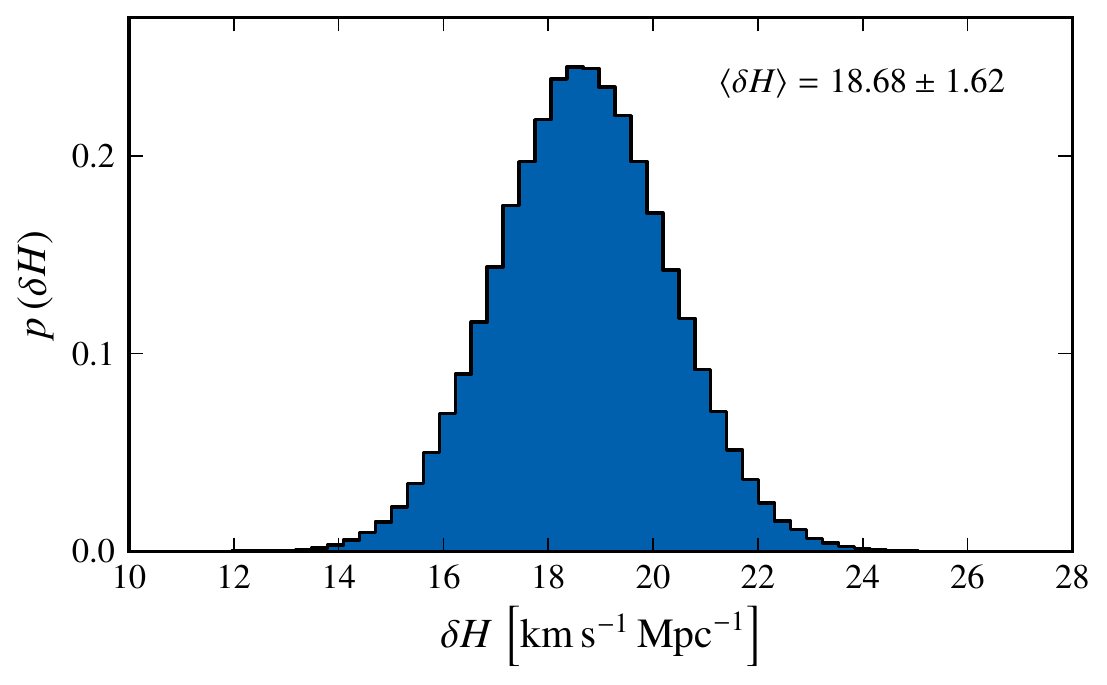}
	\caption[Average increase of the local Hubble rate due to Gpc-scale under-densities.]{Marginalised posterior distribution of the increase of the local Hubble rate $H_{\mathrm{LS}}$ (Eq. \eqref{eq:local_hubble_rate}) at the origin (due to the under-density) with respect to the Hubble rate $H_{\mathrm{bg}}$ of the homogeneous background model, $\delta H = H_{\mathrm{LS}} - H_{\mathrm{bg}}$. LTB models with fixed radial size $L = 3 \, \mathrm{Gpc}$ and a local matter density profile represented by cubic splines with three equidistant interpolation nodes were fitted to observational data of the local Hubble rate and supernovae. The LTB models were asymptotically embedded into an Einstein-de-Sitter background.}
	\label{fig:SNe_Hubble_dH}
\end{figure}

The average increase of the expansion rate around the central observer due to the large-scale under-density is the last aspect worth being considered in more detail. We therefore defined $\delta H = H_{\mathrm{LS}} - H_{\mathrm{bg}}$, which measures the difference between the effective local Hubble rate $H_{\mathrm{LS}}$ (cf. Eq.~\ref{eq:local_hubble_rate}) at the void centre and the global Hubble rate $H_{\mathrm{bg}}$ of the background model. From the mean value of the physical matter density $\omega_{\mathrm{m}}$ in Fig.~\ref{fig:SNe_Hubble_triangle}, we extracted that the data favour  $H_{\mathrm{bg}} \approx 56 \, \mathrm{km} \, \mathrm{s}^{-1} \, \mathrm{Mpc}^{-1}$, which requires an average shift of $\delta H \approx 18 \, \mathrm{km} \, \mathrm{s}^{-1} \, \mathrm{Mpc}^{-1}$ to fulfil the observationally measured Hubble rate of $H_0 \approx 74 \, \mathrm{km} \, \mathrm{s}^{-1} \, \mathrm{Mpc}^{-1}$. The magnitude of this shift agrees well with previous works \citep{2007JCAP...02..019E}. The marginalised posterior distribution of $\delta H$ is plotted in Fig.~\ref{fig:SNe_Hubble_dH}. Even though the voids can become extremely deep, the local Hubble rate rarely increases by more than $\sim 24 \, \mathrm{km} \, \mathrm{s}^{-1} \, \mathrm{Mpc}^{-1}$. This tendency is also important for Sect.~\ref{sec:SNe_Hubble_CMB}.

\subsection{Constraints: Cosmic microwave background}
\label{sec:CMB}

We proceed by analysing the implications of only the model-independent CMB constraints from Table~\ref{tab:CMB_constraints}. As reference, we first ran an MCMC simulation to fit curved FLRW models with vanishing cosmological constant to the data. These models can fit the CMB constraints just as well as flat FLRW models with non-vanishing cosmological constant. The data favour closed FLRW models with dimensionless Hubble parameter $h = 0.33 \pm 0.01$, matter density parameter $\Omega_{\mathrm{m}} = 1.26 \pm 0.04$, and consequently a curvature parameter of $\Omega_{\mathrm{k}} \approx -0.26$. It is important to note that although curved FLRW models can easily fit the minimal CMB constraints without a cosmological constant, the required Hubble rate is extremely low and strongly contradicts the expansion rate measured by \citet{2011ApJ...730..119R}.

Next, we compared the same LTB models as in the previous section ($L = 3 \, \mathrm{Gpc}$; three equidistant interpolation nodes) with the CMB data. This time, however, we asymptotically embedded the LTB models into curved FLRW backgrounds. Of course, these cosmological models fit the data perfectly as well. The constraints on the physical matter density $\omega_{\mathrm{m}} = \omega_{\mathrm{b}} + \omega_{\mathrm{c}}$ and the dimensionless Hubble parameter $h$ of the background model are essentially identical to those found without an LTB inhomogeneity around the observer. Again, an extremely low expansion rate of $h = 0.33 \pm 0.01$ is favoured. The shape of the local matter density profile is not constrained at all. The posteriors of the densities $\rho(r_i)$ at the interpolation nodes are constant over the whole prior ranges, indicating that the Monte Carlo walkers can vary the shape of the local matter density profile without notably deteriorating the fit to the CMB. Moreover, the best-fitting LTB models show almost arbitrarily shaped local density profiles. These results are the same for LTB models with different radial sizes, more or less interpolation nodes, or density profiles represented by Laguerre polynomials.

We conclude that the CMB data alone only constrain the global properties of the background FLRW model. In particular, good fits without a cosmological constant require an extremely low Hubble parameter of $h < 0.4$. The exact functional form of the local curvature and the matter density profiles are essentially irrelevant. These results agree well with the work of \citet{2009JCAP...07..029C}, who found that the small angle fluctuations of the CMB spectrum only constrain spatial curvature near the surface of last scattering.

\subsection{Constraints: $H_0$ + supernovae + cosmic microwave
background}
\label{sec:SNe_Hubble_CMB}

Finally, we simultaneously fitted LTB models to the local Hubble rate, supernovae, and the CMB data. Before that, however, we fitted spatially flat FLRW models with non-vanishing cosmological constant to the data. The best-fitting model has the cosmological parameters $ \left( h, \Omega_{m}, \Omega_{\Lambda} \right) = (0.719, 0.265, 0.735)$ and a log-likelihood value of $\log \left( \mathcal{L} \right) = -273.45$. This model -- and in particular its log-likelihood value -- serves as a reference point for the following discussion.

\begin{figure}
	\centering
	\includegraphics[width=0.45\textwidth]{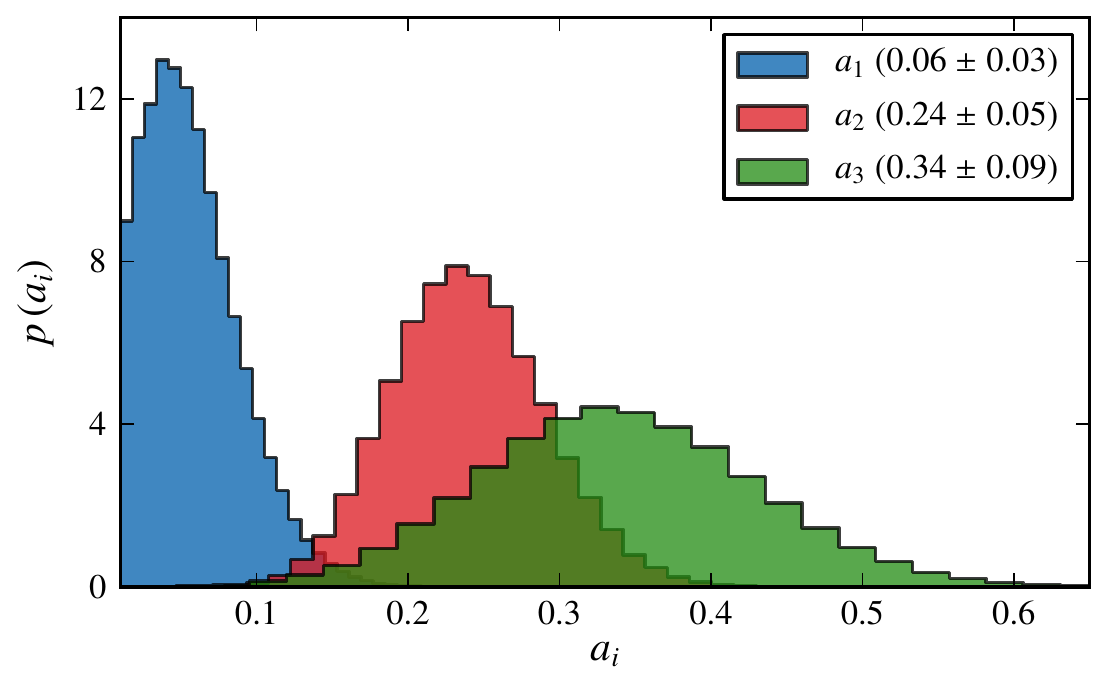}
	\caption[Constraints on the linearly interpolated density profile of LTB models fitted to $H_0$, supernovae and the CMB.]{Constraints on the linearly interpolated matter density profile of LTB models with fixed radial size $L = 3 \, \mathrm{Gpc}$ and three equidistant interpolation nodes at $r_1 = 0$, $r_2 = 1 \, \mathrm{Gpc}$ and $r_3 = 2 \, \mathrm{Gpc}$. The coefficients $a_i$ denote the matter densities at these nodes normalised with respect to the background density today, viz. $a_i = \rho(t_0, r_i)/\rho_{\mathrm{bg}}(t_0)$. The LTB models were constrained by observational data of the local Hubble rate, supernovae, and the CMB.}
	\label{fig:CMB_SNe_Hubble_linear}
\end{figure}

\begin{figure}
	\centering
	\includegraphics[width=0.45\textwidth]{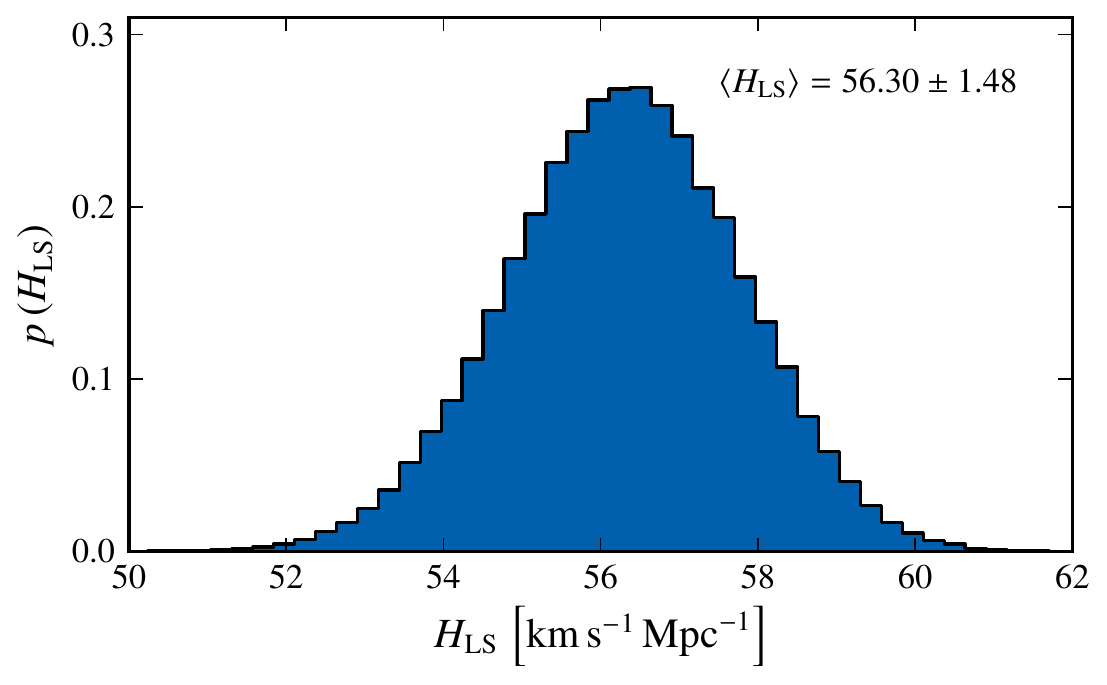}
	\caption[Effective local Hubble rate of LTB models with linear matter density profile constrained by $H_0$, supernovae and the CMB.]{Marginalised posterior distribution of the effective local Hubble rate $H_{\mathrm{LS}}$ (Eq. \eqref{eq:local_hubble_rate}) of LTB models with radial size $L = 3 \, \mathrm{Gpc}$ and linearly interpolated (using three equidistant nodes) matter density profile. The models were fitted to observational data of the local Hubble rate, supernovae, and the CMB.}
	\label{fig:CMB_SNe_Hubble_H_LS}
\end{figure}

We now considered LTB models with fixed radial size $L = 3 \, \mathrm{Gpc}$ and linearly interpolated matter density profile with three equidistant interpolation nodes at the radial coordinates  $r_1 = 0$, $r_2 = 1 \, \mathrm{Gpc}$ and $r_3 = 2 \, \mathrm{Gpc}$. For these models, the data favour an FLRW background with dimensionless Hubble parameter $h = 0.39 \pm 0.01$ and matter density parameter $\Omega_{m} = 1.09 \pm 0.03$, demonstrating that curved FLRW backgrounds can indeed improve the fit to the CMB. The statistical constraints on the local matter density profile at the present time $t_0$ are shown in Fig.~\ref{fig:CMB_SNe_Hubble_linear}. While the overall shape of the favoured profiles agrees well with our previous results from Sect.~\ref{sec:SNe_Hubble}, the voids become noticeably deeper when including the constraints from the CMB. This change becomes most apparent for the matter density at the origin, $a_1 = \rho \left( t_0, r = 0 \right)/\rho_{\mathrm{bg}(t_0)}$, indicating that the best agreement with the data can be achieved with an almost vacuum solution around the observer at the coordinate centre.

This tendency demonstrates the problem of LTB models with a constant bang time function we described above: A reasonable fit to the CMB --most importantly the angular diameter distance to the LSS -- requires an unrealistically low background Hubble parameter of $H_0 \approx 39 \, \mathrm{km} \, \mathrm{s}^{-1} \, \mathrm{Mpc}^{-1}$. On the other hand, the LTB models also need to comply with the observationally measured, high local Hubble rate of $H_0 \approx 73.8 \, \mathrm{km} \, \mathrm{s}^{-1} \, \mathrm{Mpc}^{-1}$. The Monte Carlo walkers therefore minimise the matter density at the origin to maximise the local expansion rate. However, Fig.~\ref{fig:CMB_SNe_Hubble_H_LS} shows that even though the voids can become extremely deep, the effective local Hubble rate never exceeds $\sim 62 \, \mathrm{km} \, \mathrm{s}^{-1} \, \mathrm{Mpc}^{-1}$. The best-fitting model, for instance, has the profile parameters $(a_1, a_2, a_3) = (0.04, 0.22, 0.32)$, which means that the matter density almost vanishes at the origin. The local Hubble rate still is much too low, $H_{\mathrm{LS}} = 56.83 \, \mathrm{km} \, \mathrm{s}^{-1} \, \mathrm{Mpc}^{-1}$, however, and the log-likelihood value of $\log \left( \mathcal{L} \right) = -318.55$ indicates that the model clearly cannot compete with simple, spatially flat FLRW models with cosmological constant.

As discussed in Sect.~\ref{sec:data_H0}, we computed the effective local Hubble rate $H_{\mathrm{LS}}$ by means of a least-squares fit to the Taylor-expanded luminosity distance for small redshifts, $z < 0.1$. Legitimately, one might therefore ask whether a radially fine-tuned matter density profile in the nearby range ($0 < z(r) < 0.1$) might be capable of solving the above tension. We refute this idea by adding one additional interpolation node at $r = 500 \, \mathrm{Mpc}$ and re-running the fitting procedure. The best-fitting model improves only marginally, with a slightly higher effective local Hubble rate of   $H_{\mathrm{LS}} = 57.26 \, \mathrm{km} \, \mathrm{s}^{-1} \, \mathrm{Mpc}^{-1}$ and a still inacceptable log-likelihood value of $\log \left( \mathcal{L} \right) = -317.30$. Consequently, we note that even radial fine-tuning of the matter density profile cannot mitigate the tension between the required high local Hubble rate and a good fit to the CMB.

\begin{table}
\caption[Best-fitting LTB models with varying radial size constrained by $H_0$, supernovae and the CMB.]{Evolution of the best-fitting LTB models for varying radial size $L$ in Gpc. $h$ and $\Omega_{\mathrm{m}}$ denote the dimensionless Hubble parameter and the matter density parameter of the background FLRW model, respectively. The matter density profiles were parametrised by cubic spline interpolation schemes with three equidistant nodes at the radial coordinates $r_i = (i-1) \times L/3$. The coefficients $a_i$ denote the matter densities at these nodes normalised with respect to the background density today, viz. $a_i = \rho(t_0, r_i)/\rho_{\mathrm{bg}}(t_0)$. $H_{\mathrm{LS}}$ is the effective Hubble rate measured at the origin. The last column shows the log likelihood values. The LTB models fitted to observational data of the local Hubble rate, supernovae, and the CMB.}
\label{tab:CMB_SNe_Hubble_cubic_BF}
\centering
\begin{tabular}{c|c|c|c|c|c|c|c}
$L$ & $h$ & $\Omega_{\mathrm{m}}$ & $a_1$ & $a_2$ & $a_3$ & $H_{\mathrm{LS}}$ & $\log \left( \mathcal{L} \right)$ \\
\hline \Tstrut
3 & 0.39 & 1.10 & 0.07 & 0.27 & 0.45 & 56.01 & -320.14 \\
\hline \Tstrut
4.5 & 0.39 & 1.18 & 0.01 & 0.12 & 0.17 & 60.13 & -299.95 \\
\hline \Tstrut
7.5 & 0.39 & 1.37 & 0.01 & 0.12 & 0.44 & 62.52 & -293.26
\end{tabular}
\end{table}

Lastly, we discuss the impact of the radial void size on the fit to the data. We considered LTB models whose density profiles are represented by cubic spline interpolations with three equidistant nodes at the radial coordinates $r_i = (i-1) \times L/3$. Table~\ref{tab:CMB_SNe_Hubble_cubic_BF} summarises the best-fitting models for ascending radial sizes, $L = 3 \, \mathrm{Gpc}$, $L = 4.5 \, \mathrm{Gpc}$ and $L = 7.5 \, \mathrm{Gpc}$. Obviously, the fit to the data significantly improves as the voids become larger. The favoured scenarios are extremely deep Gpc-scale under-densities, which are asymptotically embedded into increasingly dense FLRW backgrounds.

This antipodal behaviour is caused by the constraints from the CMB; the background FLRW models need to increase the convergence to decrease the angular diameter distance to the LSS. Again, even though the best-fitting models are essentially empty at the origin, the effective local Hubble rate is clearly too low compared with the observed value. We thus note that even gigantic voids of radial size $7.5 \, \mathrm{Gpc}$ with radially fine-tuned matter density profile are inconsistent with current data. Clearly, these models cannot compete with the standard cosmological model, which has a log-likelihood value of $\log \left( \mathcal{L} \right) = -273.45$.

We could continue this procedure and construct ever larger voids with more free interpolation nodes. Indeed, while testing our code, we observed that LTB voids with radial sizes $\gtrsim 12 \; \mathrm{Gpc}$ slowly converge to the same log-likelihood value as the standard cosmological model. It does not make much sense,
however, to seriously consider even larger voids than those presented in Table~\ref{tab:CMB_SNe_Hubble_cubic_BF} as faithful representations of the observable Universe for mainly two reasons. Firstly, these models become yet more implausible with growing size. We recall that we envisaged voids that emerged from fluctuations of the primordial matter distribution. For these voids to reach a size of several Gpc today, the primordial under-densities must have been extreme. These scenarios are therefore highly unlikely within the standard inflationary paradigm \citep[see also][]{2013PhRvL.110x1305M}. Secondly, as was shown for example by \citet{2008JCAP...09..016G}, \citet{2011CQGra..28p4005Z} or \citet{2012PhRvD..85b4002B}, large voids generate a pronounced kSZ effect that is strongly inconsistent with current observational data.

The results of this section agree qualitatively well with previous works, which consistently found that LTB models that fit the CMB data exhibit unrealistically low local Hubble rates \cite[e.g.][]{2010JCAP...11..030B, 2011PhRvD..83j3515M, 2012PhRvD..85b4002B}. We now seek a theoretical explanation for this empirical result.

\section{Theoretical arguments for considering $\Lambda$LTB models}
\label{sec:theoretical_explanation}

\subsection{Why vacuum solutions maximise the local Hubble rate}

In the previous section, we found that huge Gpc-scale voids with almost constant vacuum solution around the origin are favoured when the local Hubble rate and the CMB data are to be fitted simultaneously. Intuitively, it is comprehensible that such vacuum solutions indeed maximise the local Hubble rate. However, we can also understand this result using Raychaudhuri's equation \citep{1955PhRv...98.1123R}, which in LTB models is

\begin{equation}
\label{eq:raychaudhuri_ltb}
\dot{H}_{\mathrm{L}} + 2 \dot{H}_{\mathrm{T}} = - \frac{1}{3} \theta^2 - \sigma^2 - 4 \pi G \, \rho \; , 
\end{equation}
with expansion $\theta^2 = \left( H_{\mathrm{L}} + 2 H_{\mathrm{T}} \right)^2$, shear $\sigma^2 = \frac{2}{3} \left( H_{\mathrm{L}} - H_{\mathrm{T}} \right)^2$, and a dot indicating the derivative with respect to an appropriately chosen affine parameter. Generally, the shear in LTB models vanishes as $r \to 0$, implying that the longitudinal and transversal Hubble rates become identical close to the origin \citep{2006igrc.book.....P}. For Gpc-scale voids with almost constant density profile, however, the approximation $H_{\mathrm{L}} \approx H_{\mathrm{T}}$ is accurate even for larger radii, in particular in the redshift interval ($z < 0.1$) in which we determine the effective local Hubble rate (cf. Sect.~\ref{sec:data_H0}). In the region around a central observer, LTB solutions are then practically indistinguishable from simple FLRW models, and Eq. \eqref{eq:raychaudhuri_ltb} simplifies to

\begin{equation}
\label{eq:raychaudhuri_ltb_origin}
3 \dot{H} = -3 H^2 - 4 \pi G \, \rho \; .
\end{equation}
Equation \eqref{eq:raychaudhuri_ltb_origin} describes the time evolution of the Hubble rate as a function of the matter density. Starting from an initial Hubble rate $H_{\mathrm{i}}$ at an early time $t_{\mathrm{i}}$ (e.g. a time $t_{\mathrm{i}}$ at which the Universe was still homogeneous), we can readily integrate Eq. \eqref{eq:raychaudhuri_ltb_origin} to compute the expansion rate at a later time $t_{\mathrm{f}}$
\begin{equation}
\label{eq:raychauduri_ltb_origin_sol}
H(t_{\mathrm{f}}) = \dfrac{1}{\frac{1}{H_{\mathrm{i}}} + \int_{t_{\mathrm{i}}}^{t_{\mathrm{f}}} \left( 1 + \frac{1}{2} \Omega_m(t') \right) \; \mathrm{d} t'} \; ,
\end{equation}
where we introduced the common matter density parameter  $\Omega_{\mathrm{m}} = \frac{8 \pi G \rho}{3 H^2}$. Clearly, the final Hubble rate is maximised by minimising the denominator of Eq. \eqref{eq:raychauduri_ltb_origin_sol}, which is achieved by setting $\Omega_{\mathrm{m}}$ to zero. The local expansion rate of Gpc-scale voids with slowly varying density profile is thus indeed maximised by vacuum solutions.

\subsection{Why not even vacuum LTB solutions exhibit a sufficiently high local Hubble rate}

\begin{figure}
	\centering
	\includegraphics[width=0.45\textwidth]{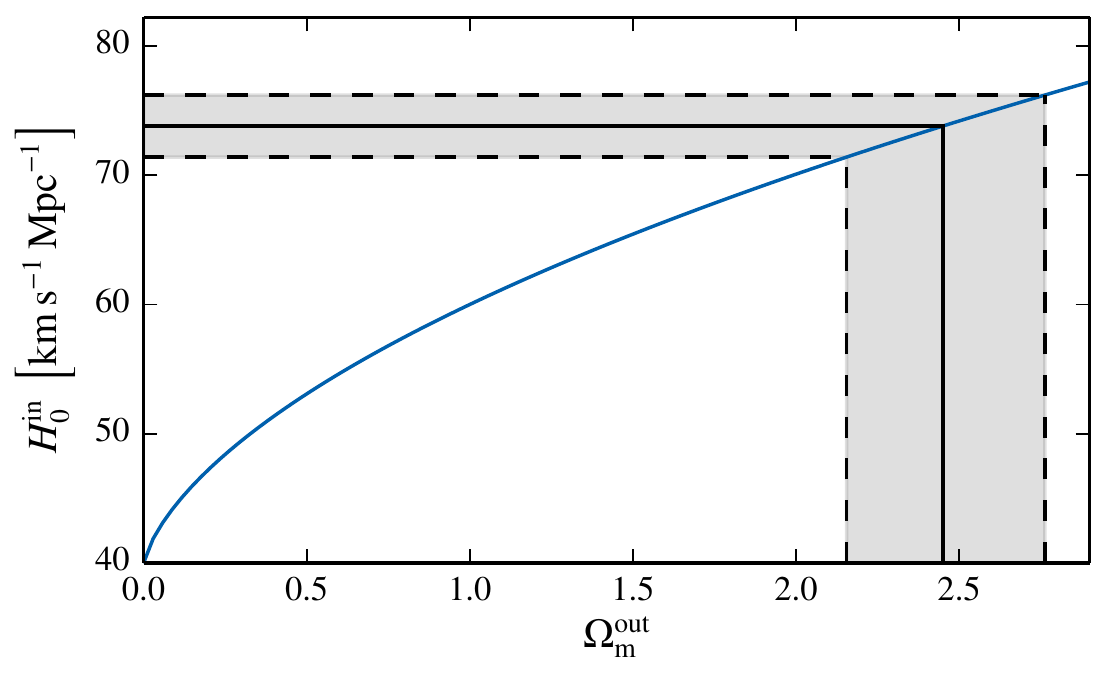}
	\caption[Hubble rate inside the empty sphere as a function of background matter density.]{Hubble rate $H_0^{\mathrm{in}}$ inside an empty sphere around the observer as a function of the matter density parameter $\Omega_{\mathrm{m}}^{\mathrm{out}}$ of the background FLRW model. The black solid line and the grey-shaded area indicate the local Hubble rate and the corresponding 68\% confidence interval as measured by \citet{2011ApJ...730..119R}.}
	\label{fig:H_in_omega}
\end{figure}

Given the result of the previous section, we can now introduce a vast simplification that helps us explain why not even empty LTB models (that simultaneously fulfil the model-independent constraints from the CMB) exhibit an effective local Hubble rate that comes close to the observationally measured value of ${H_0 \approx 73.8 \, \mathrm{km} \, \mathrm{s}^{-1} \, \mathrm{Mpc}^{-1}}$. We recall that we work in the synchronous time gauge and with a constant bang time function, meaning that the Universe has the same global age $t_0$ everywhere. We can now envisage a simplified top-hat scenario in which the Universe is globally described by a curved FLRW model, but the observer is located at the centre of an empty sphere ($\Omega_{\mathrm{m}} = 0$, i.e. a Milne model) that was carved out of the homogeneous background. For the purpose of this simplified scenario, technical details concerning the embedding or smooth junction conditions are not relevant. The Hubble function in both regions is given by
\begin{equation}
\label{eq:flrw_hubble_func}
H^2(a) = H_0^2 \left( \Omega_{\mathrm{m}} a^{-3} + \Omega_{\mathrm{k}} a^{-2} \right) \; ,
\end{equation}
where $a$ denotes the usual scale factor (with $a(t_0) = 0$) and the curvature parameter is determined by $\Omega_{\mathrm{k}} = 1 - \Omega_{\mathrm{m}}$. Integrating Eq. \eqref{eq:flrw_hubble_func} in time yields the standard FLRW relation for the age of the Universe
\begin{equation}
\label{eq:flrw_age_int}
t_0 = \frac{1}{H_0} \int_0^1 \frac{1}{\sqrt{\Omega_{\mathrm{m}} a^{-1} + \Omega_{\mathrm{k}}}} \, \mathrm{d}a \; .
\end{equation}
As this age is assumed to be globally the same, we can directly relate the Hubble rate $H_0^{\mathrm{in}}$ inside the empty sphere surrounding the observer to the Hubble rate $H_0^{\mathrm{out}}$ of the curved background,
\begin{equation}
\label{eq:h_in_h_out}
H_0^{\mathrm{in}} = \frac{H_0^{\mathrm{out}}}{f \left(\Omega_{\mathrm{m}}^{\mathrm{out}}\right)} \; , \; f \left(\Omega_{\mathrm{m}}^{\mathrm{out}}\right) = \int_0^1 \frac{1}{\sqrt{\Omega_{\mathrm{m}}^{\mathrm{out}} a^{-1} + \left( 1- \Omega_{\mathrm{m}}^{\mathrm{out}} \right)}} \; .
\end{equation}
Figure~\ref{fig:H_in_omega} shows the resulting relation between the Hubble rate $H_0^{\mathrm{in}}$ and the matter density parameter $\Omega_{\mathrm{m}}^{\mathrm{out}}$ for a fixed background Hubble rate of ${H_0^{\mathrm{out}} = 40 \, \mathrm{km} \, \mathrm{s}^{-1} \, \mathrm{Mpc}^{-1}}$, which is the value that was favoured by the CMB data in Sect.~\ref{sec:SNe_Hubble_CMB}. For instance, if we embed the empty sphere into an Einstein-de Sitter (EdS) background, we find $f \left( \Omega_{\mathrm{m}}^{\mathrm{out}} = 1 \right) = 2/3$, such that the Hubble rate at the observer position becomes ${H_0^{\mathrm{in}} = 60 \, \mathrm{km} \, \mathrm{s}^{-1} \, \mathrm{Mpc}^{-1}}$. This agrees well with several previous works, which empirically found that large and deep voids that are embedded into EdS backgrounds cannot increase the local Hubble rate by more than ${\sim 20 \, \mathrm{km} \, \mathrm{s}^{-1} \, \mathrm{Mpc}^{-1}}$ \citep{2011CQGra..28p4004M}. Our simple approximation also agrees surprisingly well with the results from Table~\ref{tab:CMB_SNe_Hubble_cubic_BF}. Most importantly, however, Fig.~\ref{fig:H_in_omega} shows that unrealistically high values for the matter density parameter $\Omega_{\mathrm{m}}^{\mathrm{out}}$ are required to reach the observationally measured local Hubble rate. These theoretical solutions are excluded by current observational data (e.g. constraints from the CMB). 

We can thus safely conclude that not even giant LTB voids with radial sizes of several Gpc and vanishing matter density at the origin can simultaneously fit the observed local Hubble rate and the CMB. 

\subsection{Inhomogeneous Big Bang or cosmological constant?}
\label{sec:inhomogeneous_big_bang}

The decisive property that we exploited in the previous section to link the local Hubble rate around the observer to the Hubble rate of the background model was the assumption of a constant global age throughout the whole Universe. The problem of a too low local Hubble rate can easily be solved by dropping this assumption. To see this very clearly, we used some trivial substitutions to rewrite Eq. \eqref{eq:LTB_age} in the standard form
\begin{equation}
t - t_{\mathrm{B}}(r) = \frac{1}{H_0(r)} \int_0^1 \frac{1}{\sqrt{ \Omega_{\mathrm{m}}(r) a^{-3} + \Omega_{\mathrm{k}}(r) a^{-2} + \Omega_{\mathrm{\Lambda}}  }} \; \mathrm{d}a \; ,
\end{equation}
where the density parameters $\Omega_{x}(r)$ can have an explicit radial dependence in LTB models. Obviously, the local Hubble rate can be increased by decreasing the age of the local Universe. Varying bang time functions that increase towards the origin ($r = 0$) are thus yet another mechanism to mimic the apparent acceleration of the Universe. In fact, \citet{2014PhRvD..89b3520K} showed that the bang time function can be calibrated such that LTB models exactly reproduce the distance-redshift relation of the standard cosmological model (with dark energy).

These models have previously been discussed in the literature \citep{2009JCAP...07..029C, 2012PhRvD..85b4002B}. They were shown to solve the tension between the local Hubble rate and the CMB data. However, \citet{2012PhRvD..85b4002B} demonstrated that models with the required fluctuations in the bang time function produce a pronounced kSZ effect in galaxy clusters, which is strongly inconsistent with current data.

In addition, we have fundamental objections against LTB models with both flexible density profile and fine-tuned bang time function. As we explained in Sect.~\ref{sec:results_ltb_zero_lambda}, the tension between the considered LTB models and current observational data arises as soon as constraints from the CMB are taken into account. In Sect. \eqref{sec:LTB_metric}, we briefly discussed that variations in the bang time function can be identified with decaying modes in linear perturbation theory, which in turn would imply (strong) inhomogeneities at early times. However, such features strongly contradict the standard CMB analysis, which is inherently based on the assumption of a spatially isotropic and homogeneous early Universe. We do not believe that it makes sense to introduce a new feature -- the varying bang time function -- only to reconcile our theoretical model with the CMB, while at the same time this new feature violates the basic assumptions the CMB analysis is based upon. Furthermore, from a statistical point of view, LTB models with varying bang time function become yet more complex, with additional degrees of freedom. Given current observational data, Ockham's razor would clearly favour the standard cosmological model.

These complications can be partially avoided by extending the considered LTB models in a more natural way. \citeauthor{1971JMP....12..498L} (\citeyear{1971JMP....12..498L, 1972JMP....13..874L}) proved two remarkable, but often overlooked theorems concerning the uniqueness of Einstein's field equations. Among other things, he showed that under very general simplicity conditions and in four dimensions, any metric theory of gravity locally conserving energy-momentum must have two coupling constants, and its metric must satisfy Einstein's field equations,
\begin{equation}
G_{\mu \nu} = \kappa T_{\mu \nu} + \Lambda g_{\mu \nu} \; ,
\end{equation}
where $G_{\mu \nu}$, $T_{\mu \nu}$ and $g_{\mu \nu}$ are, respectively, the Einstein tensor, the energy-momentum tensor, and the metric tensor. $\kappa$ and $\Lambda$ are the two coupling constants, which we identify with Newton's gravitational constant and with the cosmological constant.

According to Lovelock's theorems, the cosmological constant $\Lambda$ thus appears naturally not only in Einstein's field equations, but also in the field equations of any other metric theory of gravity. It could (and perhaps should) therefore be interpreted as a second coupling constant of the metric to matter, on a par with Newton's gravitational constant. Numerous alternative cosmological models -- such as LTB voids -- have been constructed to avoid the cosmological constant or dark energy. Nonetheless, from the point of view of Lovelock's theorem, such models must be considered incomplete unless they justify why the cosmological constant should vanish.

Following this line of reasoning, we believe that considering LTB models with non-zero cosmological constant is the most natural extension of our approach. This is an interesting step because it slightly shifts the research focus away from the dark energy problem to more general tests of the Copernican principle. For example, using flexible parametrisations of the local matter density profile, $\Lambda$LTB models allow us to derive statistical constraints on possible deviations from spatial homogeneity. In addition, effects of varying spatial curvature along the past-null cone can be explored. Finally, we can investigate whether fluctuations of the local matter density profile can noticeably influence the statistical inference of the best-fitting parameters of the standard cosmological model \citep[see e.g.][]{2013MNRAS.431.1891M, 2014MNRAS.438L...6V}. We briefly discuss these applications of $\Lambda$LTB models in the next section.

\section{Probing spatial homogeneity with $\Lambda$LTB models}
\label{sec:results_lltb}

In this section, we constrain the cosmological models considered by all data sets mentioned in Sect.~\ref{sec:obs_data}, that
is, we compute the likelihood given the local Hubble rate, supernovae, model-independent CMB constraints and kSZ data. Again, as a point of reference, we first detail the goodness-of-fit of the standard cosmological model, which is usually assumed to be a spatially flat FLRW model with non-zero cosmological constant. The best-fitting model with cosmological parameters $\left( h, \Omega_{\mathrm{m}}, \Omega_{\Lambda} \right) = \left(0.72, 0.263, 0.737 \right)$ exhibits a log-likelihood value of $\log \left( \mathcal{L} \right ) = -275.1$, with the individual contributions $\log \left[ \mathcal{L}_{H_0}(1) \right] \approx -0.3$, $\log \left[ \mathcal{L}_{\mathrm{SNe}}(580) \right] \approx -272.9$, $\log \left[ \mathcal{L}_{\mathrm{CMB}}(3) \right] \approx -0.3$ and $\log \left[ \mathcal{L}_{\mathrm{kSZ}}(9) \right] \approx -1.6$, where we explicitly indicated the number of fitted data points in parenthesis. It is remarkable how well the standard cosmological model agrees with these data. The goodness-of-fit, which can be quantified in terms of the chi-squared per degree of freedom, is truly impressive and hard to improve.

On the other hand, it is well-known that -- even within the framework of the standard cosmological model -- there seems to be a slight tension concerning the Hubble rate: while measurements with Cepheid-calibrated supernovae yield a local Hubble rate of ${H_0 = \left(73.8 \pm 2.4 \right) \, \mathrm{km} \, \mathrm{s}^{-1} \, \mathrm{Mpc}^{-1}}$, the Planck CMB data alone favour a value of ${H_0 = \left( 67.3 \pm 1.2 \right) \, \mathrm{km} \, \mathrm{s}^{-1} \, \mathrm{Mpc}^{-1}}$ \citep{2011ApJ...730..119R, 2013arXiv1303.5076P}. This discrepancy is still debated, but it could be taken as a further motivation to explore more complicated, radially inhomogeneous $\Lambda$LTB models with spatially varying Hubble rates \citep{2013PhRvL.110x1305M}. Moreover, we can address questions concerning the assumption of spatial homogeneity, such as:

\begin{itemize}

\item Can inhomogeneous cosmological models fit the given data even more accurately?

\item What limits do current data impose on fluctuations of the local matter density profile? How strongly can we deviate in our assumptions from spatial homogeneity? 

\item How do the constraints on dark energy or spatial curvature change if we drop the assumption of spatial homogeneity and marginalise over all possible radial inhomogeneities?

\end{itemize}

As discussed in Sect.~\ref{sec:CMB}, the CMB data do not constrain the detailed shape of the local matter density profile, but primarily the global properties of the background model and the spatial curvature close to the LSS. The kSZ data constrain the allowed depth of radial inhomogeneities, but since we assumed large errors for the individual measurements, current constraints are rather weak, at least when considering only moderate deviations from homogeneity. We would therefore expect that fine-tuned local matter density profiles can mainly improve the fit to the local Hubble rate and supernovae. As the constraining power of supernovae decreases with increasing redshift (cf. Sect.~\ref{sec:SNe_Hubble}), it only makes sense to consider fluctuations well inside the redshift range $z < 1.5$.

For these reasons, and also for a comparison with the previous sections, we now exemplarily consider $\Lambda$LTB models whose density profile is represented by cubic splines with three flexible nodes and a fixed radial size of $L = 3 \, \mathrm{Gpc}$. While testing, we also tried to treat the radial size $L$ as a free parameter and vary it with the Monte Carlo sampler, but the observational constraints are too weak so that the Monte Carlo chains did not converge. We asymptotically embedded these $\Lambda$LTB models into curved FLRW backgrounds.

In summary, the considered $\Lambda$LTB models are described by the cosmological parameters $h$, $\Omega_{\mathrm{m}}$, $\Omega_{\Lambda}$, $\Omega_{\mathrm{k}} = 1 - \Omega_{\mathrm{m}} - \Omega_{\Lambda}$ of the background, and the three spline nodes $\left( a_1, a_2, a_3 \right)$ parametrising possible fluctuations of the matter density profile. In contrast to the flat FLRW model, we hence introduced one new degree of freedom for the cosmological background ($\Omega_{\Lambda}$ free instead of setting $\Omega_{\Lambda} = 1 - \Omega_{\mathrm{m}}$) and three additional parameters for radial inhomogeneities. This way, our framework for testing the cosmological principle is most general and able to parametrise a broad class of space-time geometries.

\begin{figure}
	\centering
	\includegraphics[width=0.45\textwidth]{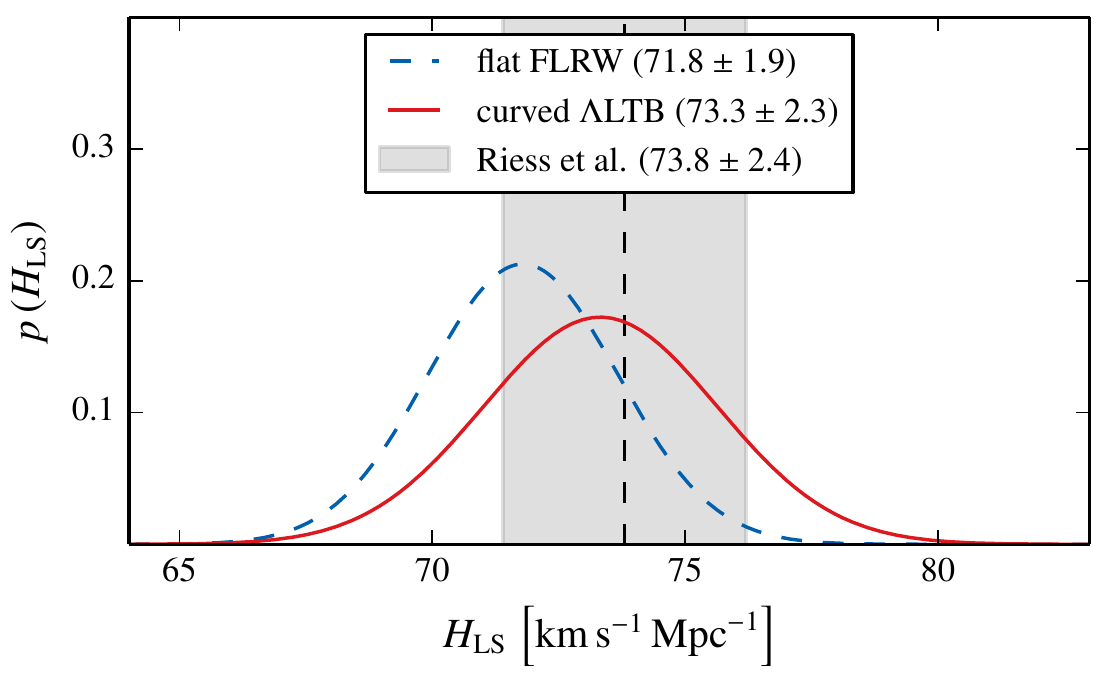}
	\caption[Marginalised posterior distributions of effective local Hubble rates.]{Marginalised posterior distributions of the effective locale Hubble rate as measured by a central observer. The blue dashed curve indicates the constraints assuming a flat FLRW model. The red curve shows the constraints assuming a $\Lambda$LTB model, with radial inhomogeneities in the range $r < 3 \, \mathrm{Gpc}$, asymptotically embedded into a curved FLRW background. The grey-shaded band indicates the local Hubble rate ($\pm 1 \sigma$) measured by \citet{2011ApJ...730..119R}. The models were constrained by data from the local Hubble rate, supernovae, the CMB, and kSZ clusters.}
	\label{fig:H_LS_FLRW_LLTB}
\end{figure}

We now sequentially discuss these questions. The best-fitting $\Lambda$LTB model has a log-likelihood value of $\log \left( \mathcal{L} \right) = -274.7$, which is only marginally better than that of the FLRW model. The fits to all observables slightly improve. There is no distinguished data set that is described significantly better.

As an example, we show a comparison of the marginalised posterior distribution of the effective local Hubble rates in Fig.~\ref{fig:H_LS_FLRW_LLTB}. As expected, the more flexible curved $\Lambda$LTB models fit the observed value better than the flat FLRW models. 

More surprisingly, the best-fitting model is almost perfectly homogeneous, with spline nodes $a_1 = 0.99$, $a_2 = 1.01$ and $a_3 = 0.99$. The cosmological parameters are also very similar to those of the standard model, with $h = 0.73$, $\Omega_{\mathrm{m}} = 0.25$, $\Omega_{\mathrm{\Lambda}} = 0.74$ and $\Omega_{\mathrm{k}} = 0.01$.

In summary, we can conclude that, of course, the more flexible models fit the data even better. The improvement is almost negligible $\left( \Delta \log \left( \mathcal{L} \right) \sim 0.4 \right)$ , however, and comes at the high price of introducing four additional free parameters. Occam's razor penalises more complicated models and can be approximated for instance by the Akaike information criterion or the Bayesian information criterion \citep{1974ITAC...19..716A, schwarz1978}. In our case, these criteria suggest that not even one additional free parameter would be justified. More meaningfully, we can thus conclude that the data used in this work statistically favour the standard cosmological model.

\begin{figure}
	\centering
	\includegraphics[width=0.45\textwidth]{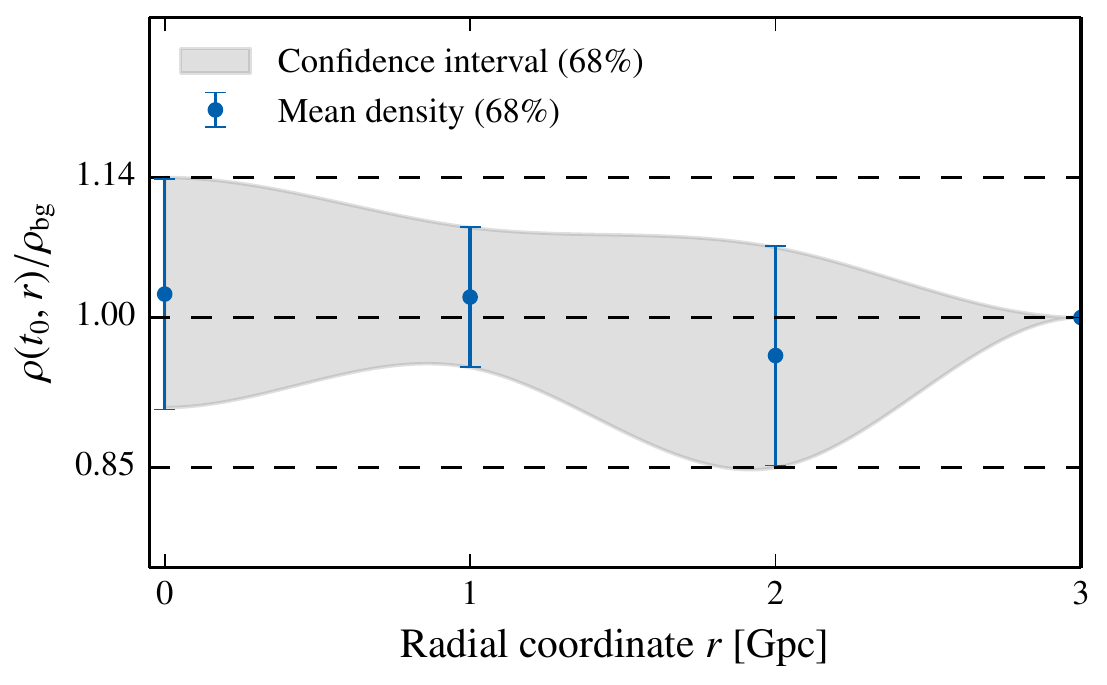}
	\caption[Statistical constraints on deviations from spatial homogeneity.]{Statistical constraints on deviations from spatial homogeneity on radial scales $r < 3 \, \mathrm{Gpc}$. The fluctuations in the matter density profile were modelled in terms of cubic splines with three equidistant nodes at $r_1 = 0$, $r_2 = 1 \, \mathrm{Gpc}$ and $r_3 = 2 \, \mathrm{Gpc}$. Constraints approaching ${r = 3 \, \mathrm{Gpc}}$ should be considered artificial, since the models are forced to converge to the background density at this radius. The models were constrained by data from the local Hubble rate, supernovae, the CMB, and kSZ clusters.}
	\label{fig:SNe_Hubble_CMB_kSZ_LLTB_3Gpc}
\end{figure}

The second question concerns statistical constraints on deviations from spatial homogeneity. The mean values of the spline parameters are given by $a_1 = 1.02 \pm 0.12$, $a_2 = 1.02 \pm 0.07$ and $a_3 = 0.96 \pm 0.11$. The corresponding variations of the density profiles are depicted in Fig.~\ref{fig:SNe_Hubble_CMB_kSZ_LLTB_3Gpc}. The density profiles were forced to converge to the background value at $r = 3 \, \mathrm{Gpc}$ by construction, so the apparent constraints at $r > 2 \, \mathrm{Gpc}$ are artificial. Clearly, the results agree well with the assumption of spatial homogeneity. However, the scatter of $\sim 15 \, \%$ indicates that we need more accurate data to safely confirm the cosmological principle. This demonstrates the importance of fully solving linear perturbation theory in LTB backgrounds (such that it can be applied to predict cosmological observables), since this would allow us to include more cosmological observables.

For instance, it would be natural to expect that Gpc-scale variations of the matter density profile leave a characteristic imprint on the large-scale structure of the local Universe. Such imprints could create non-vanishing amplitudes of the galaxy-galaxy correlation function on Gpc-scales, which would be at odds with data from current spectroscopic surveys \citep{labini2011inhomogeneities, 2012MNRAS.425..116S}. Furthermore, Gpc-variations of the local matter distribution should be well measurable with tomographic weak-lensing methods \citep{2012MNRAS.423.3445S}. There are other promising observables related to cosmic structures that might help constrain deviations from spatial homogeneity, but we need to advance the numerical algorithms for solving the linear perturbation equations on LTB backgrounds before we are able to reliably calculate these phenomena.

\begin{figure}
	\centering
	\includegraphics[width=0.45\textwidth]{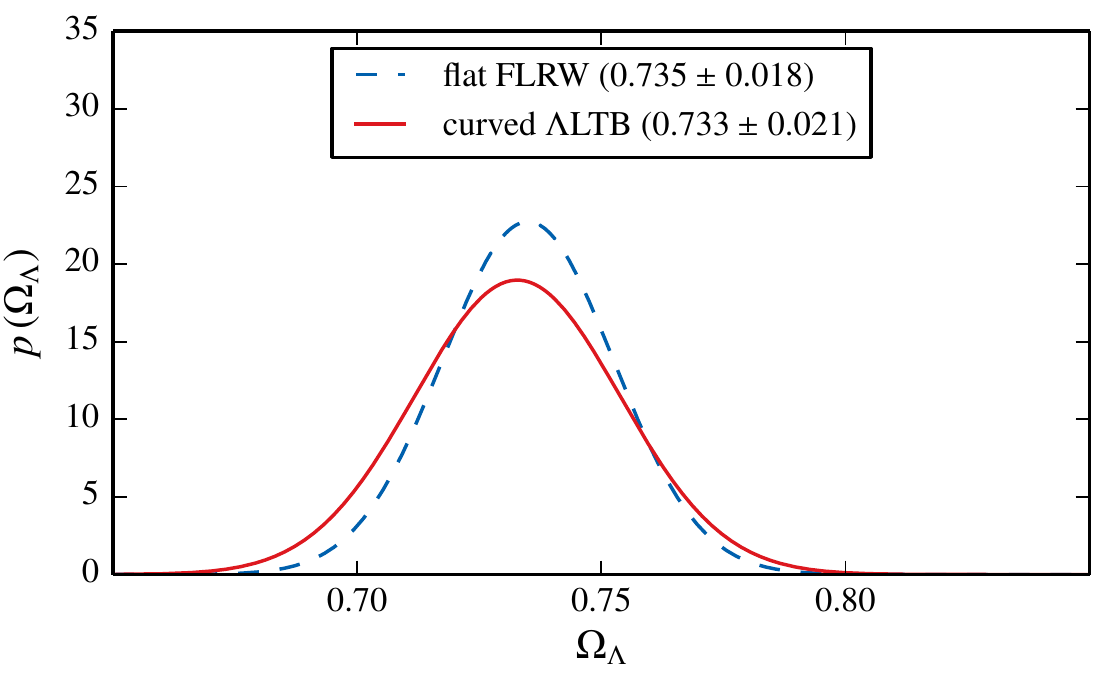}
	\caption[Marginalised posterior distributions of $\Omega_{\Lambda}$.]{Marginalised posterior distributions of the energy contribution due to the cosmological constant. The blue dashed curve indicates the constraints assuming a flat FLRW model. The red curve shows the constraints assuming a $\Lambda$LTB model, with radial inhomogeneities in the range $r < 3 \, \mathrm{Gpc}$, asymptotically embedded into a curved FLRW background. The models were constrained by data from the local Hubble rate, supernovae, the CMB, and kSZ clusters.}
	\label{fig:lambda_FLRW_LLTB}
\end{figure}

\begin{figure}
	\centering
	\includegraphics[width=0.45\textwidth]{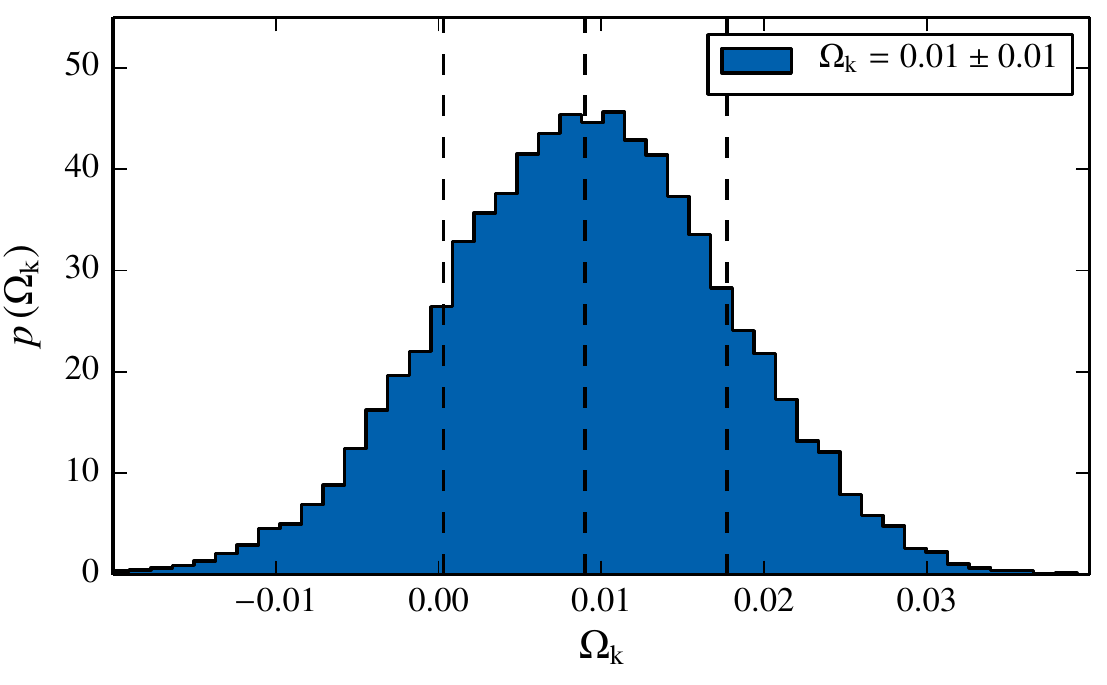}
	\caption[Marginalised posterior distribution of $\Omega_{\mathrm{k}}$.]{Posterior distribution of the spatial curvature of the background model, marginalised over fluctuations in the matter density profile at radii ${r < 3 \, \mathrm{Gpc}}$. The black dashed lines indicate the mean value and the corresponding $\pm 1 \sigma$ boundaries. The models were constrained by data from the local Hubble rate, supernovae, the CMB, and kSZ clusters.}
	\label{fig:LLTB_omega_k}
\end{figure}

To answer the last question, in Figs.  \ref{fig:lambda_FLRW_LLTB} and \ref{fig:LLTB_omega_k} we show the statistical constraints on the cosmological constant and spatial curvature. It is important to stress that these posterior distributions were marginalised over all possible fluctuations of the matter density profile. In other words, these posteriors show constraints that are independent of the assumption of spatial homogeneity, at least within the limits of our framework. The constraints on the cosmological constant are essentially invariant, with a negligibly larger scatter. This result differs from the findings of \citet{2014MNRAS.438L...6V}, who calculated a shift of $\Delta \Omega_{\Lambda} \sim 0.05$ when marginalising over inhomogeneities. However, these authors constrained their models by different data, chose another approach for modelling spatial variations of the density profile and additionally included a so-called Copernican prior. Amongst other details, these are important issues that could well explain the difference with respect to our results. Finally, it is remarkable how tightly the spatial curvature of the background model is constrained, even though we considerably relaxed our prior assumptions. The constraints are consistent with a flat background Universe, which agrees very well with the Planck results \citep{2013arXiv1303.5076P}.

\section{Conclusions}
\label{sec:conclusions}

What do cosmological observations tell us about the shape of the local matter density profile? Can current data confirm the cosmological principle? These were two central questions that we discussed in this work. To address these problems, we modelled the Universe around us by spherically symmetric, but radially inhomogeneous LTB models with a synchronous Big Bang. In contrast to most previous studies, we did not model the mass profile of LTB models in terms of empirically motivated functional forms, but instead chose to represent the local matter density profile by flexible interpolation schemes or a decomposition into Laguerre polynomials. We chose this alternative, more flexible approach to be able to investigate a broader class of problems (e.g. not restricted to void scenarios) and, more importantly, to not bias our results by prior model assumptions. 

In the first part of this work, we reconsidered LTB models without cosmological constant to investigate whether highly flexible, radially fine-tuned mass profiles allow us to simultaneously fit the high local Hubble rate and the CMB data from the Planck satellite \citep{2011ApJ...730..119R, 2013arXiv1303.5076P}. To this end, we consecutively compared numerous LTB models with different combinations of observational data. The main results of this first part can be summarised as follows:

\begin{enumerate}

\item \textbf{$\boldsymbol{H_0}$ + supernovae}: As was already well-known, LTB models without cosmological constant can easily fit the observed local Hubble rate and supernovae just as well as the standard cosmological model. These models mimic the apparent acceleration of the Universe by means of large Gpc-scale voids whose matter density profile gradually decreases towards the observer. Having said that, we would like to stress an important issue concerning the interpretation of supernova observations. Interpreted in the framework of a \textit{spatially homogeneous} FLRW model, supernovae favour a non-zero cosmological constant. Similarly, supernovae favour Gpc-scale LTB voids if and only if we set the cosmological constant to zero and require a synchronous Big Bang. In fact, \citet{2010A&A...518A..21C} demonstrated that local over-densities (i.e. giant local humps) are favoured if the bang time function is allowed to vary. We mention these ambiguities only to emphasise that the interpretation of supernovae is strongly biased by prior assumptions about the underlying cosmological model.

\item \textbf{Cosmic microwave background}: The CMB data alone do not constrain the shape of the local matter density profile. Indeed, the density profiles can substantially be varied without notably deteriorating the fit to the data. However, current CMB data impose tight constraints on the geometry and expansion rate of the asymptotic background models. Without a cosmological constant, good fits to the data require asymptotically curved FLRW backgrounds with an unrealistically low Hubble rate of ${H_0 \approx 33 \, \mathrm{km} \, \mathrm{s}^{-1} \, \mathrm{Mpc}^{-1}}$.

\item \textbf{$\boldsymbol{H_0}$ + supernovae + CMB}: The separate CMB analysis clearly highlighted the problem of the considered class of LTB models: A good fit to the CMB requires an extremely low background Hubble rate, which is in strong tension with the observed local Hubble rate of ${H_0 = (73.8 \pm 2.4) \, \mathrm{km} \, \mathrm{s}^{-1} \, \mathrm{Mpc}^{-1}}$ \citep{2011ApJ...730..119R}. In the last part of our analysis, we thus focused on the question whether or not radially tuned matter density profiles can be found that simultaneously comply with all observational constraints. However, even though we considered a wide variety of huge, extremely deep and heavily fine-tuned void profiles, the effective local Hubble rate remained too low, ${H_0 \lesssim 62 \, \mathrm{km} \, \mathrm{s}^{-1} \, \mathrm{Mpc}^{-1}}$. To make matters worse, we note that we merely used minimal, model-independent constraints from the CMB data to arrive at these conclusions (see Sect.~\ref{sec:CMB_data}, for details). We did not even use the complete information contained in the full CMB spectra, which appear to be even more problematic for void models \citep{2011PhRvD..83j3515M}. 

\end{enumerate}

After this detailed analysis, we presented simple theoretical arguments that explain why not even heavy fine-tuning of the radial matter density profile can solve the tension between the local Hubble rate and the CMB. Various solutions, such as varying bang time functions, dynamical effects of radiation, or modifications of the primordial curvature power-spectrum, were proposed to reconcile giant void scenarios with current data \citep{2011JCAP...02..013C, 2011PhRvD..83f3506N, 2012PhRvD..85b4002B}. However, all these modifications require deviations from the standard inflationary paradigm or introduce more complexity, resulting in fine-tuned, alternative cosmological models. Ockham's razor would clearly favour the standard cosmological model. In addition, according to Lovelock's theorems, the cosmological constant appears naturally in the field equations of general relativity and of any other metric theory of gravity \citep{1971JMP....12..498L, 1972JMP....13..874L}. We thus argued that considering LTB models with non-zero cosmological constant seems to be the most natural step.

In the final part, we therefore analysed LTB models with non-zero cosmological constant, which can be considered as the simplest, spatially isotropic, but radially inhomogeneous extension of the standard cosmological model. $\Lambda$LTB models are a valuable framework to systematically study deviations from spatial homogeneity, verify or falsify the cosmological principle, or simply explore effects of varying curvature along the past null cone. We showed that current data provide no evidence for radial inhomogeneities on Gpc-scales. Instead, spatially flat FLRW models with homogeneous matter distribution are favoured. These results statistically support the cosmological principle. However, we also showed that fluctuations of $\sim 15 \%$ with respect to a homogeneous matter density profile are still compatible with current data.

As emphasised in the main text (cf. Sects. \ref{sec:obs_data} and \ref{sec:results_lltb}), our analysis was limited by an important drawback: linear perturbation theory in LTB models is substantially more complicated than in FLRW models, mainly because scalar, vectorial, and tensorial perturbations do not decouple on inhomogeneous backgrounds. Although great progress has been made, substantial additional efforts are required before linear perturbations can reliably be computed in realistic cosmological settings. We therefore had to neglect all cosmological observables that depend on the details of linear structure formation. For instance, we were not yet able to calculate statistical properties of the perturbed matter density in the LTB models considered, meaning that we could not yet compare these models with the statistical properties of the observed large-scale structure (e.g. galaxy-galaxy correlation functions, and cluster number counts). Such tests should certainly help to constrain the underlying space-time geometry. We also
had to neglect important cosmological probes such as baryonic acoustic oscillations and weak-lensing spectra, which are widely (and successfully) used to constrain homogeneous and isotropic cosmologies. We will focus our research on advancing linear perturbation theory in LTB models to tighten observational constraints on the shape of the local matter density profile in future works.

\begin{acknowledgements}

MR thanks the Sydney Institute for Astronomy (SIfA, University of Sydney) for hospitality and the German Academic Exchange Service (DAAD, Doktoranden-Stipendium) for the financial support during a research visit. This work was supported in part by the \emph{Deut\-sche For\-schungs\-ge\-mein\-schaft\/} (DFG). Most simulations required for this work were performed on the bwGRiD cluster (http://www.bw-grid.de), member of the German D-Grid initiative, funded by the Ministry for Education and Research (Bundesministerium für Bildung und Forschung) and the Ministry for Science, Research and Arts Baden-Wuerttemberg (Ministerium für Wissenschaft, Forschung und Kunst Baden-Württemberg).

\end{acknowledgements}

\bibliographystyle{aa}
\bibliography{LTB}

\begin{thebibliography}{98}
\expandafter\ifx\csname natexlab\endcsname\relax\def\natexlab#1{#1}\fi

\bibitem[{Ahnert \& Mulansky(2011)}]{odeint_paper}
Ahnert, K. \& Mulansky, M. 2011, AIP Conference Proceedings, 1389, 1586

\bibitem[{{Akaike}(1974)}]{1974ITAC...19..716A}
{Akaike}, H. 1974, IEEE Transactions on Automatic Control, 19, 716

\bibitem[{{Akeret} {et~al.}(2013){Akeret}, {Seehars}, {Amara}, {Refregier}, \&
  {Csillaghy}}]{2013A&C.....2...27A}
{Akeret}, J., {Seehars}, S., {Amara}, A., {Refregier}, A., \& {Csillaghy}, A.
  2013, Astronomy and Computing, 2, 27

\bibitem[{Akima(1970)}]{Akima:1970:NMI:321607.321609}
Akima, H. 1970, J. ACM, 17, 589

\bibitem[{Alefeld {et~al.}(1995)Alefeld, Potra, \&
  Shi}]{Alefeld:1995:AEZ:210089.210111}
Alefeld, G.~E., Potra, F.~A., \& Shi, Y. 1995, ACM Trans. Math. Softw., 21, 327

\bibitem[{{Allison} \& {Dunkley}(2014)}]{2014MNRAS.437.3918A}
{Allison}, R. \& {Dunkley}, J. 2014, \mnras, 437, 3918

\bibitem[{{Alnes} \& {Amarzguioui}(2006)}]{2006PhRvD..74j3520A}
{Alnes}, H. \& {Amarzguioui}, M. 2006, \prd, 74, 103520

\bibitem[{{Amanullah} {et~al.}(2010){Amanullah}, {Lidman}, {Rubin}, {Aldering},
  {Astier}, {Barbary}, {Burns}, {Conley}, {Dawson}, {Deustua}, {Doi}, {Fabbro},
  {Faccioli}, {Fakhouri}, {Folatelli}, {Fruchter}, {Furusawa}, {Garavini},
  {Goldhaber}, {Goobar}, {Groom}, {Hook}, {Howell}, {Kashikawa}, {Kim}, {Knop},
  {Kowalski}, {Linder}, {Meyers}, {Morokuma}, {Nobili}, {Nordin}, {Nugent},
  {{\"O}stman}, {Pain}, {Panagia}, {Perlmutter}, {Raux}, {Ruiz-Lapuente},
  {Spadafora}, {Strovink}, {Suzuki}, {Wang}, {Wood-Vasey}, {Yasuda}, \&
  {Supernova Cosmology Project}}]{2010ApJ...716..712A}
{Amanullah}, R., {Lidman}, C., {Rubin}, D., {et~al.} 2010, \apj, 716, 712

\bibitem[{{Amendola} \& {Tsujikawa}(2010)}]{2010deto.book.....A}
{Amendola}, L. \& {Tsujikawa}, S. 2010, {Dark Energy: Theory and Observations}
  (Cambridge University Press)

\bibitem[{{Audren}(2013)}]{2013arXiv1312.5696A}
{Audren}, B. 2013, arXiv:1312.5696

\bibitem[{{Audren} {et~al.}(2013{\natexlab{a}}){Audren}, {Lesgourgues},
  {Benabed}, \& {Prunet}}]{2013JCAP...02..001A}
{Audren}, B., {Lesgourgues}, J., {Benabed}, K., \& {Prunet}, S.
  2013{\natexlab{a}}, \jcap, 2, 1

\bibitem[{{Audren} {et~al.}(2013{\natexlab{b}}){Audren}, {Lesgourgues},
  {Benabed}, \& {Prunet}}]{2013ascl.soft07002A}
{Audren}, B., {Lesgourgues}, J., {Benabed}, K., \& {Prunet}, S.
  2013{\natexlab{b}}, {Monte Python: Monte Carlo Code for CLASS in Python},
  astrophysics Source Code Library

\bibitem[{{Bennett} {et~al.}(2013){Bennett}, {Larson}, {Weiland}, {Jarosik},
  {Hinshaw}, {Odegard}, {Smith}, {Hill}, {Gold}, {Halpern}, {Komatsu}, {Nolta},
  {Page}, {Spergel}, {Wollack}, {Dunkley}, {Kogut}, {Limon}, {Meyer}, {Tucker},
  \& {Wright}}]{2013ApJS..208...20B}
{Bennett}, C.~L., {Larson}, D., {Weiland}, J.~L., {et~al.} 2013, \apjs, 208, 20

\bibitem[{{Biswas} {et~al.}(2010){Biswas}, {Notari}, \&
  {Valkenburg}}]{2010JCAP...11..030B}
{Biswas}, T., {Notari}, A., \& {Valkenburg}, W. 2010, \jcap, 11, 30

\bibitem[{{Bolejko}(2009)}]{2009GReGr..41.1737B}
{Bolejko}, K. 2009, General Relativity and Gravitation, 41, 1737

\bibitem[{{Bolejko} {et~al.}(2011){Bolejko}, {C{\'e}l{\'e}rier}, \&
  {Krasi{\'n}ski}}]{2011CQGra..28p4002B}
{Bolejko}, K., {C{\'e}l{\'e}rier}, M.-N., \& {Krasi{\'n}ski}, A. 2011,
  Classical and Quantum Gravity, 28, 164002

\bibitem[{{Bolejko} \& {Wyithe}(2009)}]{2009JCAP...02..020B}
{Bolejko}, K. \& {Wyithe}, J.~S.~B. 2009, \jcap, 2, 20

\bibitem[{{Bondi}(1947)}]{1947MNRAS.107..410B}
{Bondi}, H. 1947, \mnras, 107, 410

\bibitem[{{Bridle} {et~al.}(2002){Bridle}, {Crittenden}, {Melchiorri},
  {Hobson}, {Kneissl}, \& {Lasenby}}]{2002MNRAS.335.1193B}
{Bridle}, S.~L., {Crittenden}, R., {Melchiorri}, A., {et~al.} 2002, \mnras,
  335, 1193

\bibitem[{{Bull} {et~al.}(2012){Bull}, {Clifton}, \&
  {Ferreira}}]{2012PhRvD..85b4002B}
{Bull}, P., {Clifton}, T., \& {Ferreira}, P.~G. 2012, \prd, 85, 024002

\bibitem[{Carlson(1995)}]{carlson1995}
Carlson, B. 1995, Numerical Algorithms, 10, 13

\bibitem[{{C{\'e}l{\'e}rier}(2000)}]{2000A&A...353...63C}
{C{\'e}l{\'e}rier}, M.-N. 2000, \aap, 353, 63

\bibitem[{{C{\'e}l{\'e}rier}(2012)}]{2012A&A...543A..71C}
{C{\'e}l{\'e}rier}, M.-N. 2012, \aap, 543, A71

\bibitem[{{C{\'e}l{\'e}rier} {et~al.}(2010){C{\'e}l{\'e}rier}, {Bolejko}, \&
  {Krasi{\'n}ski}}]{2010A&A...518A..21C}
{C{\'e}l{\'e}rier}, M.-N., {Bolejko}, K., \& {Krasi{\'n}ski}, A. 2010, \aap,
  518, A21

\bibitem[{{Clarkson}(2012)}]{2012CRPhy..13..682C}
{Clarkson}, C. 2012, Comptes Rendus Physique, 13, 682

\bibitem[{{Clarkson} {et~al.}(2009){Clarkson}, {Clifton}, \&
  {February}}]{2009JCAP...06..025C}
{Clarkson}, C., {Clifton}, T., \& {February}, S. 2009, \jcap, 6, 25

\bibitem[{{Clarkson} \& {Regis}(2011)}]{2011JCAP...02..013C}
{Clarkson}, C. \& {Regis}, M. 2011, \jcap, 2, 13

\bibitem[{{Clifton} {et~al.}(2009){Clifton}, {Ferreira}, \&
  {Zuntz}}]{2009JCAP...07..029C}
{Clifton}, T., {Ferreira}, P.~G., \& {Zuntz}, J. 2009, \jcap, 7, 29

\bibitem[{{Clowes} {et~al.}(2013){Clowes}, {Harris}, {Raghunathan},
  {Campusano}, {S{\"o}chting}, \& {Graham}}]{2013MNRAS.429.2910C}
{Clowes}, R.~G., {Harris}, K.~A., {Raghunathan}, S., {et~al.} 2013, \mnras,
  429, 2910

\bibitem[{{Durrer} \& {Maartens}(2008)}]{2008GReGr..40..301D}
{Durrer}, R. \& {Maartens}, R. 2008, General Relativity and Gravitation, 40,
  301

\bibitem[{{Efstathiou}(2014)}]{2014MNRAS.440.1138E}
{Efstathiou}, G. 2014, \mnras, 440, 1138

\bibitem[{{Einasto} {et~al.}(2011{\natexlab{a}}){Einasto}, {Liivam{\"a}gi},
  {Tago}, {Saar}, {Tempel}, {Einasto}, {Mart{\'{\i}}nez}, \&
  {Hein{\"a}m{\"a}ki}}]{2011A&A...532A...5E}
{Einasto}, M., {Liivam{\"a}gi}, L.~J., {Tago}, E., {et~al.} 2011{\natexlab{a}},
  \aap, 532, A5

\bibitem[{{Einasto} {et~al.}(2011{\natexlab{b}}){Einasto}, {Liivam{\"a}gi},
  {Tempel}, {Saar}, {Tago}, {Einasto}, {Enkvist}, {Einasto}, {Mart{\'{\i}}nez},
  {Hein{\"a}m{\"a}ki}, \& {Nurmi}}]{2011ApJ...736...51E}
{Einasto}, M., {Liivam{\"a}gi}, L.~J., {Tempel}, E., {et~al.}
  2011{\natexlab{b}}, \apj, 736, 51

\bibitem[{{Ellis}(2009)}]{2009GReGr..41..581E}
{Ellis}, G.~F.~R. 2009, General Relativity and Gravitation, 41, 581

\bibitem[{{Enqvist}(2008)}]{2008GReGr..40..451E}
{Enqvist}, K. 2008, General Relativity and Gravitation, 40, 451

\bibitem[{{Enqvist} \& {Mattsson}(2007)}]{2007JCAP...02..019E}
{Enqvist}, K. \& {Mattsson}, T. 2007, \jcap, 2, 19

\bibitem[{{Etherington}(1933)}]{1933PMag...15..761E}
{Etherington}, I.~M.~H. 1933, Philosophical Magazine, 15, 761

\bibitem[{{Etherington}(2007)}]{2007GReGr..39.1055E}
{Etherington}, I.~M.~H. 2007, General Relativity and Gravitation, 39, 1055

\bibitem[{{February} {et~al.}(2013){February}, {Larena}, {Clarkson}, \&
  {Pollney}}]{2013arXiv1311.5241F}
{February}, S., {Larena}, J., {Clarkson}, C., \& {Pollney}, D. 2013,
  arXiv:1311.5241

\bibitem[{{Foreman} {et~al.}(2010){Foreman}, {Moss}, {Zibin}, \&
  {Scott}}]{2010PhRvD..82j3532F}
{Foreman}, S., {Moss}, A., {Zibin}, J.~P., \& {Scott}, D. 2010, \prd, 82,
  103532

\bibitem[{{Foreman-Mackey} {et~al.}(2013){Foreman-Mackey}, {Hogg}, {Lang}, \&
  {Goodman}}]{2013PASP..125..306F}
{Foreman-Mackey}, D., {Hogg}, D.~W., {Lang}, D., \& {Goodman}, J. 2013, \pasp,
  125, 306

\bibitem[{{Garcia-Bellido} \& {Haugb{\o}lle}(2008)}]{2008JCAP...04..003G}
{Garcia-Bellido}, J. \& {Haugb{\o}lle}, T. 2008, \jcap, 4, 3

\bibitem[{{Garc{\'{\i}}a-Bellido} \&
  {Haugb{\o}lle}(2008)}]{2008JCAP...09..016G}
{Garc{\'{\i}}a-Bellido}, J. \& {Haugb{\o}lle}, T. 2008, \jcap, 9, 16

\bibitem[{Goodman \& Weare(2010)}]{goodman2010ensemble}
Goodman, J. \& Weare, J. 2010, Communications in Applied Mathematics and
  Computational Science, 5, 65

\bibitem[{{Guy} {et~al.}(2007){Guy}, {Astier}, {Baumont}, {Hardin}, {Pain},
  {Regnault}, {Basa}, {Carlberg}, {Conley}, {Fabbro}, {Fouchez}, {Hook},
  {Howell}, {Perrett}, {Pritchet}, {Rich}, {Sullivan}, {Antilogus}, {Aubourg},
  {Bazin}, {Bronder}, {Filiol}, {Palanque-Delabrouille}, {Ripoche}, \&
  {Ruhlmann-Kleider}}]{2007A&A...466...11G}
{Guy}, J., {Astier}, P., {Baumont}, S., {et~al.} 2007, \aap, 466, 11

\bibitem[{Hastings(1970)}]{hastings1970monte}
Hastings, W.~K. 1970, Biometrika, 57, 97

\bibitem[{{Hellaby} \& {Krasi{\'n}ski}(2006)}]{2006PhRvD..73b3518H}
{Hellaby}, C. \& {Krasi{\'n}ski}, A. 2006, \prd, 73, 023518

\bibitem[{{Hellaby} \& {Lake}(1985)}]{1985ApJ...290..381H}
{Hellaby}, C. \& {Lake}, K. 1985, \apj, 290, 381

\bibitem[{{Hicken} {et~al.}(2009){Hicken}, {Wood-Vasey}, {Blondin}, {Challis},
  {Jha}, {Kelly}, {Rest}, \& {Kirshner}}]{2009ApJ...700.1097H}
{Hicken}, M., {Wood-Vasey}, W.~M., {Blondin}, S., {et~al.} 2009, \apj, 700,
  1097

\bibitem[{{Hogg} {et~al.}(2005){Hogg}, {Eisenstein}, {Blanton}, {Bahcall},
  {Brinkmann}, {Gunn}, \& {Schneider}}]{2005ApJ...624...54H}
{Hogg}, D.~W., {Eisenstein}, D.~J., {Blanton}, M.~R., {et~al.} 2005, \apj, 624,
  54

\bibitem[{{Hu} \& {Sugiyama}(1996)}]{1996ApJ...471..542H}
{Hu}, W. \& {Sugiyama}, N. 1996, \apj, 471, 542

\bibitem[{{Kessler} {et~al.}(2009){Kessler}, {Becker}, {Cinabro}, {Vanderplas},
  {Frieman}, {Marriner}, {Davis}, {Dilday}, {Holtzman}, {Jha}, {Lampeitl},
  {Sako}, {Smith}, {Zheng}, {Nichol}, {Bassett}, {Bender}, {Depoy}, {Doi},
  {Elson}, {Filippenko}, {Foley}, {Garnavich}, {Hopp}, {Ihara}, {Ketzeback},
  {Kollatschny}, {Konishi}, {Marshall}, {McMillan}, {Miknaitis}, {Morokuma},
  {M{\"o}rtsell}, {Pan}, {Prieto}, {Richmond}, {Riess}, {Romani}, {Schneider},
  {Sollerman}, {Takanashi}, {Tokita}, {van der Heyden}, {Wheeler}, {Yasuda}, \&
  {York}}]{2009ApJS..185...32K}
{Kessler}, R., {Becker}, A.~C., {Cinabro}, D., {et~al.} 2009, \apjs, 185, 32

\bibitem[{{Krasi{\'n}ski}(2014)}]{2014PhRvD..89b3520K}
{Krasi{\'n}ski}, A. 2014, \prd, 89, 023520

\bibitem[{{Krasi{\'n}ski} \& {Hellaby}(2002)}]{2002PhRvD..65b3501K}
{Krasi{\'n}ski}, A. \& {Hellaby}, C. 2002, \prd, 65, 023501

\bibitem[{{Labini}(2010)}]{2010AIPC.1241..981L}
{Labini}, F.~S. 2010, in American Institute of Physics Conference Series, Vol.
  1241, American Institute of Physics Conference Series, ed. J.-M. {Alimi} \&
  A.~{Fu{\"o}zfa}, 981--990

\bibitem[{Labini(2011)}]{labini2011inhomogeneities}
Labini, F.~S. 2011, Classical and Quantum Gravity, 28, 164003

\bibitem[{{Lema{\^i}tre}(1933)}]{1933ASSB...53...51L}
{Lema{\^i}tre}, G. 1933, Annales de la Societe Scietifique de Bruxelles, 53, 51

\bibitem[{{Lesgourgues}(2011)}]{2011arXiv1104.2932L}
{Lesgourgues}, J. 2011, arXiv:1104.2932

\bibitem[{{Lewis} {et~al.}(2000){Lewis}, {Challinor}, \&
  {Lasenby}}]{2000ApJ...538..473L}
{Lewis}, A., {Challinor}, A., \& {Lasenby}, A. 2000, \apj, 538, 473

\bibitem[{{Lovelock}(1971)}]{1971JMP....12..498L}
{Lovelock}, D. 1971, Journal of Mathematical Physics, 12, 498

\bibitem[{{Lovelock}(1972)}]{1972JMP....13..874L}
{Lovelock}, D. 1972, Journal of Mathematical Physics, 13, 874

\bibitem[{{Maartens}(2011)}]{2011RSPTA.369.5115M}
{Maartens}, R. 2011, Royal Society of London Philosophical Transactions Series
  A, 369, 5115

\bibitem[{{Marra} {et~al.}(2013{\natexlab{a}}){Marra}, {Amendola}, {Sawicki},
  \& {Valkenburg}}]{2013PhRvL.110x1305M}
{Marra}, V., {Amendola}, L., {Sawicki}, I., \& {Valkenburg}, W.
  2013{\natexlab{a}}, Physical Review Letters, 110, 241305

\bibitem[{{Marra} \& {Notari}(2011)}]{2011CQGra..28p4004M}
{Marra}, V. \& {Notari}, A. 2011, Classical and Quantum Gravity, 28, 164004

\bibitem[{{Marra} \& {P{\"a}{\"a}kk{\"o}nen}(2010)}]{2010JCAP...12..021M}
{Marra}, V. \& {P{\"a}{\"a}kk{\"o}nen}, M. 2010, \jcap, 12, 21

\bibitem[{{Marra} {et~al.}(2013{\natexlab{b}}){Marra}, {P{\"a}{\"a}kk{\"o}nen},
  \& {Valkenburg}}]{2013MNRAS.431.1891M}
{Marra}, V., {P{\"a}{\"a}kk{\"o}nen}, M., \& {Valkenburg}, W.
  2013{\natexlab{b}}, \mnras, 431, 1891

\bibitem[{{Melchior} {et~al.}(2014){Melchior}, {Sutter}, {Sheldon}, {Krause},
  \& {Wandelt}}]{2014MNRAS.440.2922M}
{Melchior}, P., {Sutter}, P.~M., {Sheldon}, E.~S., {Krause}, E., \& {Wandelt},
  B.~D. 2014, \mnras, 440, 2922

\bibitem[{{Moss} {et~al.}(2011){Moss}, {Zibin}, \&
  {Scott}}]{2011PhRvD..83j3515M}
{Moss}, A., {Zibin}, J.~P., \& {Scott}, D. 2011, \prd, 83, 103515

\bibitem[{{Nadathur}(2013)}]{2013MNRAS.434..398N}
{Nadathur}, S. 2013, \mnras, 434, 398

\bibitem[{{Nadathur} \& {Hotchkiss}(2014)}]{2014MNRAS.440.1248N}
{Nadathur}, S. \& {Hotchkiss}, S. 2014, \mnras, 440, 1248

\bibitem[{{Nadathur} \& {Sarkar}(2011)}]{2011PhRvD..83f3506N}
{Nadathur}, S. \& {Sarkar}, S. 2011, \prd, 83, 063506

\bibitem[{{Perlmutter} {et~al.}(1999){Perlmutter}, {Aldering}, {Goldhaber},
  {Knop}, {Nugent}, {Castro}, {Deustua}, {Fabbro}, {Goobar}, {Groom}, {Hook},
  {Kim}, {Kim}, {Lee}, {Nunes}, {Pain}, {Pennypacker}, {Quimby}, {Lidman},
  {Ellis}, {Irwin}, {McMahon}, {Ruiz-Lapuente}, {Walton}, {Schaefer}, {Boyle},
  {Filippenko}, {Matheson}, {Fruchter}, {Panagia}, {Newberg}, {Couch}, \&
  {Supernova Cosmology Project}}]{1999ApJ...517..565P}
{Perlmutter}, S., {Aldering}, G., {Goldhaber}, G., {et~al.} 1999, \apj, 517,
  565

\bibitem[{{Planck Collaboration} {et~al.}(2013){Planck Collaboration}, {Ade},
  {Aghanim}, {Armitage-Caplan}, {Arnaud}, {Ashdown}, {Atrio-Barandela},
  {Aumont}, {Baccigalupi}, {Banday}, \& et~al.}]{2013arXiv1303.5076P}
{Planck Collaboration}, {Ade}, P.~A.~R., {Aghanim}, N., {et~al.} 2013,
  arXiv:1303.5076

\bibitem[{{Plebanski} \& {Krasinski}(2006)}]{2006igrc.book.....P}
{Plebanski}, J. \& {Krasinski}, A. 2006, {An Introduction to General Relativity
  and Cosmology}

\bibitem[{{Raychaudhuri}(1955)}]{1955PhRv...98.1123R}
{Raychaudhuri}, A. 1955, Physical Review, 98, 1123

\bibitem[{{Riess} {et~al.}(1998){Riess}, {Filippenko}, {Challis},
  {Clocchiatti}, {Diercks}, {Garnavich}, {Gilliland}, {Hogan}, {Jha},
  {Kirshner}, {Leibundgut}, {Phillips}, {Reiss}, {Schmidt}, {Schommer},
  {Smith}, {Spyromilio}, {Stubbs}, {Suntzeff}, \&
  {Tonry}}]{1998AJ....116.1009R}
{Riess}, A.~G., {Filippenko}, A.~V., {Challis}, P., {et~al.} 1998, \aj, 116,
  1009

\bibitem[{{Riess} {et~al.}(2011){Riess}, {Macri}, {Casertano}, {Lampeitl},
  {Ferguson}, {Filippenko}, {Jha}, {Li}, \& {Chornock}}]{2011ApJ...730..119R}
{Riess}, A.~G., {Macri}, L., {Casertano}, S., {et~al.} 2011, \apj, 730, 119

\bibitem[{{Robertson}(1935)}]{1935ApJ....82..284R}
{Robertson}, H.~P. 1935, \apj, 82, 284

\bibitem[{{Sch{\"a}fer} \& {Heisenberg}(2012)}]{2012MNRAS.423.3445S}
{Sch{\"a}fer}, B.~M. \& {Heisenberg}, L. 2012, \mnras, 423, 3445

\bibitem[{Schwarz(1978)}]{schwarz1978}
Schwarz, G. 1978, The Annals of Statistics, 6, 461

\bibitem[{{Scrimgeour} {et~al.}(2012){Scrimgeour}, {Davis}, {Blake}, {James},
  {Poole}, {Staveley-Smith}, {Brough}, {Colless}, {Contreras}, {Couch},
  {Croom}, {Croton}, {Drinkwater}, {Forster}, {Gilbank}, {Gladders},
  {Glazebrook}, {Jelliffe}, {Jurek}, {Li}, {Madore}, {Martin}, {Pimbblet},
  {Pracy}, {Sharp}, {Wisnioski}, {Woods}, {Wyder}, \&
  {Yee}}]{2012MNRAS.425..116S}
{Scrimgeour}, M.~I., {Davis}, T., {Blake}, C., {et~al.} 2012, \mnras, 425, 116

\bibitem[{{Silk}(1977)}]{1977A&A....59...53S}
{Silk}, J. 1977, \aap, 59, 53

\bibitem[{{Smale} \& {Wiltshire}(2011)}]{2011MNRAS.413..367S}
{Smale}, P.~R. \& {Wiltshire}, D.~L. 2011, \mnras, 413, 367

\bibitem[{{Springel} {et~al.}(2005){Springel}, {White}, {Jenkins}, {Frenk},
  {Yoshida}, {Gao}, {Navarro}, {Thacker}, {Croton}, {Helly}, {Peacock}, {Cole},
  {Thomas}, {Couchman}, {Evrard}, {Colberg}, \& {Pearce}}]{2005Natur.435..629S}
{Springel}, V., {White}, S.~D.~M., {Jenkins}, A., {et~al.} 2005, \nat, 435, 629

\bibitem[{{Sunyaev} \& {Zeldovich}(1980)}]{1980MNRAS.190..413S}
{Sunyaev}, R.~A. \& {Zeldovich}, I.~B. 1980, \mnras, 190, 413

\bibitem[{{Sunyaev} \& {Zeldovich}(1970)}]{1970Ap&SS...7....3S}
{Sunyaev}, R.~A. \& {Zeldovich}, Y.~B. 1970, \apss, 7, 3

\bibitem[{{Sunyaev} \& {Zeldovich}(1972)}]{1972CoASP...4..173S}
{Sunyaev}, R.~A. \& {Zeldovich}, Y.~B. 1972, Comments on Astrophysics and Space
  Physics, 4, 173

\bibitem[{{Sutter} {et~al.}(2014){Sutter}, {Lavaux}, {Wandelt}, {Weinberg},
  {Warren}, \& {Pisani}}]{2014MNRAS.442.3127S}
{Sutter}, P.~M., {Lavaux}, G., {Wandelt}, B.~D., {et~al.} 2014, \mnras, 442,
  3127

\bibitem[{{Suzuki} {et~al.}(2012){Suzuki}, {Rubin}, {Lidman}, {Aldering},
  {Amanullah}, {Barbary}, {Barrientos}, {Botyanszki}, {Brodwin}, {Connolly},
  {Dawson}, {Dey}, {Doi}, {Donahue}, {Deustua}, {Eisenhardt}, {Ellingson},
  {Faccioli}, {Fadeyev}, {Fakhouri}, {Fruchter}, {Gilbank}, {Gladders},
  {Goldhaber}, {Gonzalez}, {Goobar}, {Gude}, {Hattori}, {Hoekstra}, {Hsiao},
  {Huang}, {Ihara}, {Jee}, {Johnston}, {Kashikawa}, {Koester}, {Konishi},
  {Kowalski}, {Linder}, {Lubin}, {Melbourne}, {Meyers}, {Morokuma}, {Munshi},
  {Mullis}, {Oda}, {Panagia}, {Perlmutter}, {Postman}, {Pritchard}, {Rhodes},
  {Ripoche}, {Rosati}, {Schlegel}, {Spadafora}, {Stanford}, {Stanishev},
  {Stern}, {Strovink}, {Takanashi}, {Tokita}, {Wagner}, {Wang}, {Yasuda},
  {Yee}, \& {Supernova Cosmology Project}}]{2012ApJ...746...85S}
{Suzuki}, N., {Rubin}, D., {Lidman}, C., {et~al.} 2012, \apj, 746, 85

\bibitem[{{Tolman}(1934)}]{1934PNAS...20..169T}
{Tolman}, R.~C. 1934, Proceedings of the National Academy of Science, 20, 169

\bibitem[{{Valkenburg}(2012)}]{2012GReGr..44.2449V}
{Valkenburg}, W. 2012, General Relativity and Gravitation, 44, 2449

\bibitem[{{Valkenburg} {et~al.}(2014){Valkenburg}, {Marra}, \&
  {Clarkson}}]{2014MNRAS.438L...6V}
{Valkenburg}, W., {Marra}, V., \& {Clarkson}, C. 2014, \mnras, 438, L6

\bibitem[{{Vonlanthen} {et~al.}(2010){Vonlanthen}, {R{\"a}s{\"a}nen}, \&
  {Durrer}}]{2010JCAP...08..023V}
{Vonlanthen}, M., {R{\"a}s{\"a}nen}, S., \& {Durrer}, R. 2010, \jcap, 8, 23

\bibitem[{{Walker}(1935)}]{1935QJMat...6...81W}
{Walker}, A.~G. 1935, The Quarterly Journal of Mathematics, 6, 81

\bibitem[{{Zibin}(2008)}]{2008PhRvD..78d3504Z}
{Zibin}, J.~P. 2008, \prd, 78, 043504

\bibitem[{{Zibin} \& {Moss}(2011)}]{2011CQGra..28p4005Z}
{Zibin}, J.~P. \& {Moss}, A. 2011, Classical and Quantum Gravity, 28, 164005

\bibitem[{{Zibin} {et~al.}(2008){Zibin}, {Moss}, \&
  {Scott}}]{2008PhRvL.101y1303Z}
{Zibin}, J.~P., {Moss}, A., \& {Scott}, D. 2008, Physical Review Letters, 101,
  251303

\bibitem[{{Zumalac{\'a}rregui} {et~al.}(2012){Zumalac{\'a}rregui},
  {Garc{\'{\i}}a-Bellido}, \& {Ruiz-Lapuente}}]{2012JCAP...10..009Z}
{Zumalac{\'a}rregui}, M., {Garc{\'{\i}}a-Bellido}, J., \& {Ruiz-Lapuente}, P.
  2012, \jcap, 10, 9

\end{thebibliography}

\begin{appendix}

\section{Parametric solutions for the LTB model without cosmological constant}
\label{app:LTB}

In this appendix, we amend some more useful relations for LTB models without cosmological constant that we skipped in the main part to improve the readability of the paper. 

To begin with, we repeat Eq. \eqref{eq:LTB_age} (with $\Lambda = 0$) from Sect.~\ref{sec:LTB_metric}, which results from integrating one of Einstein's field equations in time:

\begin{equation}
t_0 - t_{\mathrm{B}}(r) = \int\limits_0^{R(t_0,r)} \frac{1}{\sqrt{\frac{2M(r)}{\tilde{R}} + 2E(r)}} \, \mathrm{d}\tilde{R} \; .
\end{equation}
Dependent on the sign of the curvature function $E(r)$, this integral has three different parametric solutions:

\begin{itemize}

\item Elliptic evolution: $E(r) < 0$ 

	\begin{align}
	\label{eq:ps_R_e}
	R(t,r) &= -\frac{M(r)}{2E(r)}(1 - \cos\eta) \; , \\
	\eta - \sin\eta &= \frac {\left[ -2E(r)\right]^{3/2}}{M(r)} \left[ t - 			t_{\mathrm{B}}(r) \right] \; .
	\end{align}

\item Parabolic evolution: $E(r) = 0$

	\begin{equation}
	R(t,r) = \left\{ \frac{9}{2} M(r) \left[t - t_{\mathrm{B}}(r) \right]^2 \right\}^{1/3} \; .
	\end{equation}

\item Hyperbolic evolution: $E(r) > 0$

	\begin{align}
	R(t,r) &= \frac{M(r)}{2E(r)} (\cosh\eta - 1) \; , \\
	\label{eq:ps_theta_h}
	\sinh\eta - \eta &= \frac{\left[2E(r)\right]^{3/2}}{M(r)} \left[t - t_{\mathrm{B}}(r) \right] \; .
	\end{align}

\end{itemize}
As described in Sect.~\ref{sec:LTB_algorithm}, we determine the curvature function $E(r)$ as a function of time $t$, effective gravitational mass $M(r)$, and areal radius $R(t,r)$. For this purpose, it is useful to rewrite the parametric solutions as

\begin{align}
t =& \; t_{\mathrm{B}} + \frac{M}{(-2E)^{3/2}} \left[ \arccos \left( 1 + \frac{2ER}{M} \right) - 2 \sqrt{ \frac{-ER}{M} \left( 1 + \frac{ER}{M} \right) } \, \right] \nonumber \\
\label{eq:ps_t_ex}
 & 0 \leq \eta \leq \pi \; ,  \\
t =& \; t_{\mathrm{B}} + \frac{M}{(-2E)^{3/2}} \left[ \pi + \arccos \left( - 1 - \frac{2ER}{M} \right) + 2  \sqrt{ \frac{-ER}{M} \left( 1 + \frac{ER}{M} \right) } \, \right] \nonumber \; , \\
\label{eq:ps_t_ec}
& \pi \leq \eta \leq 2 \pi \; ,
\end{align}
for the expanding and collapsing elliptic cases, and

\begin{equation}
\label{eq:ps_t_hx}
t = t_{\mathrm{B}} + \frac{M}{(2 E)^{3/2}} \left[ 2\sqrt{ \frac{ER}{M} \left( 1 + \frac{ER}{M} \right) } - \mathrm{arcosh} \left( 1 + \frac{2ER}{M} \right) \right]
\end{equation}
for the hyperbolic case. We refer for example to \citet{2006igrc.book.....P} for more details on these solutions.  Equations \eqref{eq:ps_t_ex}, \eqref{eq:ps_t_ec} and \eqref{eq:ps_t_hx} can be numerically unstable in the near parabolic limit ($|E| \ll 1$). In this case, we use an inverse series expansion as explained in Appendix B of \citet{2006PhRvD..73b3518H}.

Finally, we note that the parametric solutions \eqref{eq:ps_R_e} - \eqref{eq:ps_theta_h} can be combined to derive an analytic expression for the radial derivative of the areal radius function,

\begin{equation}
\label{eq:LTB_R_prime}
{R'} = \left( \frac{M'}{M} - \frac{E'}{E} \right) R + \left[ \left(\frac{3}{2} \frac{E'}{E} - \frac{M'}{M} \right) (t-t_{\mathrm{B}}) - t_{\mathrm{B}}' \right] \dot{R} \; .
\end{equation}
This relation, together with the radial derivative of Eq. \eqref{eq:LTB_EFE1},
\begin{equation}
\dot{R}' = \frac{1}{\dot{R}} \, \left[ \left( \frac{M'R-MR'}{R^2} \right) + E'  \right] \; ,
\end{equation}
allows us to analytically compute $R'$ and $\dot{R}'$, which renders the numerical integration of radial null geodesics more efficient (cf. Eqs. \eqref{eq:LTB_dtdr} and \eqref{eq:LTB_dzdr}).

\end{appendix}

\end{document}